\documentclass[11pt]{article}  
\usepackage{graphicx}
\usepackage{amssymb}
\usepackage{color}

\input epsf
\parskip=\medskipamount
\def\al{\alpha}

   \def\Ga{\Gamma}

\def\la{\lambda}  
\def\kp{\kappa}  
\def\te{\theta}   
   
\def\IB{\relax{\rm l\kern-.18 em B}}
\def\IC{\relax{\rm l\kern-.50 em C}}
\def\IE{\relax{\rm l\kern-.12 em E}}
\def\IK{\relax{\rm l\kern-.18 em K}}
\def\IL{\relax{\rm I\kern-.18 em L}}
\def\IN{\relax{\rm I\kern-.18 em N}}
\def\IR{\relax{\rm I\kern-.18 em R}}
\def\ii{\rm i\,}

\def\\{\hfill\break}

\def\tr{\mathop{\rm tr}\nolimits}
\def\Re{\mathop{\rm Re}\nolimits}
\def\Im{\mathop{\rm Im}\nolimits}

\def\Cos{\mathop{\rm C}\nolimits}    
\def\Sin{\mathop{\rm S}\nolimits}    
\def\Tan{\mathop{\rm T}\nolimits}    
\font\tenfrak=eufm10  \font\sevenfrak=eufm7  \font\fivefrak=eufm5
\newfam\frakfam
\textfont\frakfam=\tenfrak
\scriptfont\frakfam=\sevenfrak
\scriptscriptfont\frakfam=\fivefrak
\def\frak{\fam\frakfam\tenfrak}

\newtheorem{theorem}{Theorem}
\newtheorem{proposition}{Proposition}
\newtheorem{property}{Property}

\def\frac#1#2{{#1\over #2}}
\def\fracpd#1#2{\frac{\partial #1}{\partial #2}}
\def\ptos{\leaders\hbox to 2mm{\hfil{.}\hfil}\hfill}

\begin{document}

\title{Superintegrability on the 3-dimensional  spaces with curvature.\\ 
Oscillator-related and Kepler-related systems  on  \\ the Sphere $S^3$ and on the Hyperbolic space $H^3$ }

\author{
Jos\'e F.\ Cari\~nena$^{\dagger\,a)}$,
Manuel F.\ Ra\~nada$^{\dagger\,b)}$, and
Mariano Santander$^{\ddagger\,c)}$ \\ [2pt]
$^\dagger${\sl Departamento de F\'{\i}sica Te\'orica and IUMA, Facultad de Ciencias} \\
   {\sl Universidad de Zaragoza, 50009 Zaragoza, Spain}  \\   [2pt]
$^\ddagger${\sl Departamento de F\'{\i}sica Te\'{o}rica and IMUVa, Facultad de Ciencias} \\
   {\sl Universidad de Valladolid, 47011 Valladolid, Spain} 
} 
\date{}
\maketitle 


\begin{abstract}  
The superintegrability of several Hamiltonian systems defined on three-dimensional configuration spaces of constant curvature is studied. 
We first analyze the properties of the Killing vector fields, Noether symmetries  and Noether momenta.
Then we study the superintegrability of the Harmonic Oscillator, the Smorodinsky-Winternitz (S-W) system and the Harmonic Oscillator with ratio of frequencies 1:1:2 and additional nonlinear terms on the 3-dimensional sphere $S^3$ ($\kp>0)$ and on the hyperbolic space $H^3$ ($\kp<0$).  
In the second part we present a study first of the Kepler problem and then of the Kepler problem with additional nonlinear terms in these two curved spaces, $S^3$ ($\kp>0)$ and $H^3$ ($\kp<0$).
We prove  their superintegrability and we obtain, in all the cases,  the  maximal number of functionally independent integrals of motion.  

 All the mathematical expressions are presented using the curvature $\kp$ as a parameter, in such a way that particularizing for $\kp>0$, $\kp=0$,  or $\kp<0$, the corresponding properties are obtained for the system on the sphere $S^3$, the Euclidean space $\IE^3$, or the hyperbolic space $H^3$, respectively.  

\end{abstract}

\begin{quote}
{\sl Keywords:}{\enskip} Superintegrability ;  Constant curvature spaces ;  
Oscillator-related Hamiltonians ;  Kepler-related Hamiltonians ; Constants of motion  ;  
Higher-order constants of motion 

{\sl Running title:}{\enskip}
Superintegrability of 3-dimensional  Hamiltonian systems with constant curvature.

AMS classification:  37J06 ;  37J35   ; 70H06  ;  70H33
\end{quote}

\vfill
\footnoterule{\small
\begin{quote}
$^{a)}${\it E-mail address:} {jfc@unizar.es } \\
$^{b)}${\it E-mail address:} {mfran@unizar.es }  \\
$^{c)}${\it E-mail address:} {mariano.santander@uva.es }
\end{quote}
}


 \tableofcontents

\section{Introduction}
A Hamiltonian system  is superintegrable if it is integrable in the Liouville sense  and it  admits  more globally defined  constants of motion than degrees of freedom.   If the system has three degrees of freedom and it admits five functionally independent  integrals of motion then the system is maximally superintegrable.  At the classical level superintegrability means that all bounded trajectories are closed while at the quantum level this property is related to the degeneracy of the energy levels.
The two best known examples of these systems are the harmonic oscillator and the Kepler-Coulomb problem. In fact, probably the first and oldest study on this matter was the theorem of Bertrand \cite{Bertrand,SantosBertrand} (although of course without using this word) which states that the only central potentials for which all bounded trajectories are closed are just these two particular systems: isotropic oscillator and Kepler potential. 
In these two cases the additional integrals of motion are the components of the Fradkin tensor \cite{Fradkin65} for the harmonic oscillator, and of the Runge--Lenz vector in the case of the Kepler system \cite{Goldstein75,Goldstein76}.

Fris et al \cite{FrisMSUW65} studied the superintegrability in the Euclidean plane and proved the existence of four families of superintegrable systems with quadratic in the momenta constants of motion; two of these families were related with the harmonic oscillator and the other two with the Kepler problem. 
This research was then continued in three-dimensional Euclidean space by Evans \cite{Ev90PR} and then other systems were studied in different situations as  on spaces with constant curvature \cite{Slawianowski00}--\cite{Gonera20},  on  two-dimensional pseudo-Euclidean spaces \cite{Ran97Jmp}--\cite{Camp14Jmp},  on spaces with conformally Euclidean metrics \cite{KaKM05Jmp}--\cite{CRS21Jpa} and even on more general curved spaces \cite{KaKW02}--\cite{CHR17Jmp}   (see \cite{MillPW13} for a review). 
Most of these systems were endowed with quadratic  integrals of motion but systems possessing  integrals of motion of higher-order have also been studied \cite{PopPostW12Jmp}--\cite{MarqWint19}, mainly in the two-dimensional Euclidean space. 
We also mention some other recent articles  dealing with different aspects of superintegrability \cite{Shang18}--\cite{Ghazouani20}. 

 In the present paper we analyze the superintegrability first of the oscillator and the Kepler systems and then of some oscillator-related and Kepler-related Hamiltonian systems  on  three-dimensional spaces of constant curvature, that is, on the Sphere $S^3$ and  the Hyperbolic space $H^3$. 
Actually, the paper is related to some previous papers that were also concerned with similar problems. 
Some of them were related with classical Hamiltonian systems on two-dimensional spaces \cite{CRS05Jmp, Ran14Jpa,Ran15PLa}, \cite{RS02Jmp,RS03Jmp} and others  were devoted to the quantum superintegrability on $S^2$ and $H^2$ \cite{CRS07AnnPhys,CRS11Jmp,CRS12Jmp} and also on $S^3$ and $H^3$ \cite{CRS11Ijtp,CRS12Jpa}.

It is clear that spherical and hyperbolic spaces are endowed with quite different geometrical properties 
but nevertheless  some dynamical properties (as for example those related with the integrability of Hamiltonian systems) can be studied by making use of a joint approach valid for the two types of spaces.
So, the main idea is to study at the same time in a joint or unified form (and not as two different studies) both situations: dynamics on the sphere (curvature $\kp$ positive) and dynamics on the Hyperbolic space (curvature $\kp$ negative). 
With this aim we will make use of the similar notation and techniques introduced  in the above-mentioned previous articles. 
The following  points summarize the main characteristics of this approach.

(i) All the mathematical expressions are presented using the curvature  $\kp$ as a parameter, in such a way that particularizing for   $\kp>0$, $\kp=0$,  or $\kp<0$, the corresponding properties are obtained for the system on the sphere $S^3$, the Euclidean space $\IE^3$, or the hyperbolic space $H^3$, respectively.  

(ii) The limit when $\kp\to 0$ is always well defined and when $\kp=0$, all the characteristics of the Euclidean system are recovered. 

(iii) Many different $\kp$-dependent potentials can be constructed satisfying properties (i) and (ii). 
Nevertheless,  if we require that the superintegrability must be preserved, then this condition determines  a  particular $\kp$-dependent  function among all the possible curved version of the Euclidean potential,

A consequence of this approach is that we obtain the following result

(iv)  All the fundamental properties of the superintegrable Euclidean system continue to hold for the curvature-modified Hamiltonian with  $\kp\ne 0$ (in both cases $\kp>0$ and $\kp<0$) but they appear in a modified  $\kp$-dependent way.

The main idea is that  the spherical and hyperbolic versions of the harmonic oscillator and Kepler system can be considered as deformations of the well known Euclidean systems, and conversely, the Euclidean harmonic oscillator and Kepler system are very particular cases of the more general ``curved" systems.
This fact remains also true for the noncentral systems obtaining by adding some additional nonlinear terms to these two important central potentials. 

In more detail, the plan of this article is as follows:
In Section 2 we first present the notation and then we study the existence of Killing vector fields, the Lagrangian and the Hamiltonian for the geodesic free motion, the existence of Noether symmetries and the properties of the Noether momenta. 
Then in Section 3  we  first study the harmonic oscillator on $S^3$ and $H^3$ and then an oscillator-related Hamiltonian with three nonlinear terms of the form $k_1/x^2$, $k_2/y^2$, and $k_3/z^2$. In Section 4 we study the noncentral 2:1:1 oscillator  also with additional  nonlinear terms and in Section 5 first the Kepler problem and then we analyze a Kepler system modified with the three $\kp$-dependent nonlinear terms $k_1/x^2$,  $k_2/y^2$  and $k_3/z^2$ and we prove that it  is also superintegrable but with constants of motion of fourth order in the momenta.
Finally, in Section 6 we make some final comments.

\section{Geodesic motion, $\kp$-dependent formalism, Killing vector fields, and Noether momenta }

In the following, all the mathematical expressions will depend of the curvature  $\kp$ as a parameter, in such a way that for  $\kp>0$, $\kp=0$, or $\kp<0$, we will obtain the corresponding property  of the Hamiltonian system particularized on the three-dimensional spaces Sphere $S^3$, Euclidean space $\IE^3$ and  Hyperbolic space $H^3$. 

In order to obtain appropriate curvature-dependent expressions we will make use the following curvature-dependent trigonometric and hyperbolic functions:
\begin{equation}
  \Cos_{\kp}(x) = \cases{
  \cos{\sqrt{\kp}\,x}       &if $\kp>0$, \cr
  {\quad}  1               &if $\kp=0$, \cr
  \cosh\!{\sqrt{-\kp}\,x}   &if $\kp<0$, \cr}{\qquad}
  \Sin_{\kp}(x) = \cases{
  \frac{1}{\sqrt{\kp}} \sin{\sqrt{\kp}\,x}     &if $\kp>0$, \cr
  {\quad}   x                                &if $\kp=0$, \cr
  \frac{1}{\sqrt{-\kp}}\sinh\!{\sqrt{-\kp}\,x} &if $\kp<0$, \label{SkCk}\cr}
\end{equation}
and the $\kp$-dependent tangent defined in the natural way
$$
  \Tan_{\kp}(x) = \frac{ \Sin_{\kp}(x)}{ \Cos_{\kp}(x) } ,
$$
so that the Euclidean limits are correctly defined
$$
 \lim{}_{\kp\to 0}\Cos_{\kp}(x) = 1 \,,{\quad} 
 \lim{}_{\kp\to 0}\Sin_{\kp}(x)  = x \,,{\quad} 
 \lim{}_{\kp\to 0}\Tan_{\kp}(x) = x \,,  
$$ 
(these functions, that were used in the papers   \cite{RS02Jmp}--\cite{CRS12Jpa} mentioned in the introduction, have also been used by other authors, see, e.g. \cite{Gonera14,Gonera20} and \cite{ChanuDgRast14,ChanuRast17}). 
 In this way we  can express in a single $\kp$-dependent formula  the differential element of distance $ds^2(\kp)$ in geodesic spherical coordinates $(r,\te,\phi)$ on  the three three-dimensional manifolds Sphere $S^3$, Euclidean space $\IE^3$ and   Hyperbolic space $H^3$
\begin{equation}
  ds^2(\kp) =  dr^2  +  \Sin_\kp^2(r)\,d\te^2   +  \Sin_\kp^2(r)\,\sin^2\te d\phi^2   \,,   
\end{equation}
where we recall that $r$ denotes  the geodesic distance to the origin  (North pole in the spherical case) and not the radius of a sphere. 
It reduces to 
\begin{eqnarray*}
 ds_1^2  &=&    dr^2  +  \sin^2 r\,d\te^2 +   \sin^2 r \sin^2\te\,d\phi^2    \,, \cr
 ds_0^2  &=&    dr^2  +  r^2 \,d\te^2 + r^2 \sin^2\te\,d\phi^2      \,, \cr
  ds_{-1}^2 &=&   dr^2  +  \sinh^2 r\,d\te^2 +   \sinh^2 r \sin^2\te\,d\phi^2 \,, 
\end{eqnarray*}
in the three particular cases of the unit sphere $(\kp=1)$, Euclidean space $(\kp=0)$, and 'unit'  Lobachewski space $(\kp=-1)$.

\subsection{Lagrangian formalism, Noether symmetries, and Noether momenta }

We recall that a vector field $X \in {\frak X}(Q)$, defined on a Riemann manifold $(Q,g)$,  is called a Killing vector field when it preserves the metric in the sense that the Lie derivative with respect to $X$ of the tensor metric $g$ vanishes.
A three-dimensional Riemann manifold with constant curvature  admits six linearly independent Killing vector fields, that generate six different one-parameter groups of continuous isometries of the manifold, and that can be grouped in two sets of three vector fields.
\begin{itemize}
\item[(i)]  Three $\kp$-dependent vector fields that we denote by  $X_j$, $j=1,2,3$,
$$ 
  X_1  =   (\sin{\te}\cos{\phi}) \fracpd{}{r} +  \left(\frac{\Cos_\kp(r)}{\Sin_\kp(r)} \right) 
  \bigl[(\cos{\te}\cos{\phi})\,\fracpd{}{\te} - \bigl(\frac{\sin{\phi}}{\sin\te}\bigr)\,\fracpd{}{\phi}  \bigr] \,,
$$
$$ 
  X_2  =   (\sin{\te}\sin{\phi}) \fracpd{}{r} + \left(\frac{\Cos_\kp(r)}{\Sin_\kp(r)} \right)
  \bigl[(\cos{\te}\sin{\phi})\,\fracpd{}{\te}  + \bigl( \frac{\cos{\phi}}{\sin\te}\bigr)\,\fracpd{}{\phi}  \bigr] \,,
$$
$$ 
  X_3  =   (\cos{\te})\,\fracpd{}{r} - \left(\frac{\Cos_\kp(r)}{\Sin_\kp(r)} \right) \sin{\te}\,\fracpd{}{\te}  \,,
$$
\item[(ii)]  Three  $\kp$-independent  vector fields  that we denote by  $Y_j$, $j=1,2,3$,
$$
  Y_1 =   -\sin\phi\,\fracpd{}{\te}
      - \left(\frac{\cos{\phi}}{\tan \te}\right)\,\fracpd{}{\phi} \,,{\qquad}
  Y_2 =   \cos\phi\,\fracpd{}{\te}
      - \left(\frac{\sin{\phi}}{\tan \te}\right)\,\fracpd{}{\phi} \,,{\qquad}
  Y_3 =   \fracpd{}{\phi} \,,
$$
\end{itemize}
Everyone of these six vector fields is the generator of  a  one-parameter group of diffeomorphisms preserving the metric $ds_\kp^2$, that is, a one-parameter group of isometries of the Riemannian manifold. 
 Moreover if we denote by $\Omega_\kp$ the the volume form determined by the metric $g$,
$$
 \Omega_\kp = \sqrt{|\,g\,|} \,dr \wedge d\te \wedge d\phi = 
 \Sin_\kp^2(r)\sin\te \,dr \wedge d\te \wedge d\phi \,,{\quad} |\,g\,| = \det g \,, 
$$
then they also preserve the volume form, that is, 
$$
 {\cal L}_{X_i}\Omega_\kp = 0 \,,{\quad}  {\cal L}_{Y_i}\Omega_\kp = 0  
 \,,{\quad} i=1,2,3, 
$$
where ${\cal L}_X$ denotes the Lie derivative with respect to the vector field $X$. 

 Moreover,  they close the following Lie algebra,  
$$
 [X_1\,, X_2] = -\,\kp\,Y_3 \,,{\quad}  [X_2\,, X_3] = -\,\kp\,Y_1 \,,{\quad} 
 [X_3\,, X_1] =  -\,\kp\,Y_2\,,
$$
$$
 [Y_1\,, Y_2] =  -Y_3\,,{\quad}  [Y_2\,, Y_3] =  -Y_1\,,{\quad} [Y_3\,, Y_1] = -Y_2\,,
$$
that represents the Lie algebra of the isometries of the three-dimensional spherical,  Euclidean, and hyperbolic  spaces. 
Notice that in the limit $\kp\to 0$ we recover the Euclidean algebra and  the vector fields $X_j$,  $j=1,2,3$,  commute between themselves.

The Lagrangian for the geodesic  free  motion is given by the kinetic term determined by the metric
\begin{equation}
 L (r,\theta,\phi,v_r,v_\theta,v_\phi)= T_g(\kp)  = (\frac{1}{2})\,\left(v_r^2 + \Sin_\kp^2(r)\, v_\te^2 +  \Sin_\kp^2(r)\sin^2\te \,v_\phi^2 \right) \,,     
\end{equation}
where the parameter $\kp$ can take any real value. 
Three particular cases are  $\kp=-1$, $\kp=0$, and $\kp=1$ 
\begin{eqnarray*}
  L_1(r,\theta,\phi,v_r,v_\theta,v_\phi)  &=& (\frac{1}{2})\,\left(v_r^2 +  \sin^2 r\,v_\te^2 +   \sin^2 r \sin^2\te\,v_\phi^2 \right) \,,\cr 
  L_0 (r,\theta,\phi,v_r,v_\theta,v_\phi)  &=&  (\frac{1}{2})\,\left(v_r^2 + r^2 \,v_\te^2 + r^2\sin^2\te\,v_\phi^2\right) \,, \cr
  L_{-1} (r,\theta,\phi,v_r,v_\theta,v_\phi) &=&  (\frac{1}{2})\,\left(v_r^2 +  \sinh^2 r\,v_\te^2 +   \sinh^2 r \sin^2\te\,v_\phi^2 \right) . 
\end{eqnarray*}

This kinetic Lagrangian possesses six exact Noether symmetries, because  $L = T_g(\kp)$ is invariant under the action of the six Killing vectors in the sense that we have
$$
  X_i^t(T_g(\kp) )= 0 \,,{\quad}  Y_i^t(T_g(\kp) )= 0 \,,{\quad} i=1,2,3,
$$
where $X_i^t$ and $Y_i^t$ denote the tangent lifts (or complete lifts) of $X_i$ and $Y_i$ to the velocity phase space $TQ$ where $Q$ is $Q=S^3$, $Q=\IE^3$ or $Q=H^3$,
according to the value of $\kp$. 
 
\begin{theorem}
{\rm Noether Theorem}:  Each Killing vector field is  a symmetry of the geodesic Lagrangian $L =  T_g$, and hence  it determines a constant of the `geodesic' motion. 
\end{theorem}

If we denote by $\te_L$ the Lagrangian one-form:
\begin{eqnarray*}
  \te_L &=&  \fracpd{L}{v_r} dr +  \fracpd{L}{v_\te} d\te + \fracpd{L}{v_\phi} d\phi    \cr
   &=&  v_r \,dr + \Sin_\kp^2(r) v_\te\,d\te +  \Sin_\kp^2(r)\sin^2\te \,v_\phi\,d\phi
 \end{eqnarray*}
then the associated Noether constants of the motion (that are usually known as Noether momenta) are given by the following: 

\begin{itemize}
\item[(P)]   The three $\kp$-dependent functions $P_1(\kp)$, $P_2(\kp)$, and $P_3(\kp)$, defined as
$$
 P_1 =  i(X_1)\,\te_L  \,,\quad 
 P_2 =  i(X_2)\,\te_L  \,,\quad 
 P_3 =  i(X_3)\,\te_L     \,,
$$
that are given by
\begin{eqnarray}
  P_1 &=&  (\sin{\te}\cos\phi)\,{v_r} + \bigl(\Cos_\kp(r)\Sin_\kp(r)\bigr)
   \bigl[(\cos{\te}\cos{\phi})\,v_\te - (\sin{\te}\sin{\phi})\,v_\phi\bigr]   \,, \cr
 P_2 &=&  (\sin{\te}\sin\phi)\,{v_r} + \bigl(\Cos_\kp(r)\Sin_\kp(r)\bigr)
   \bigl[(\cos{\te}\sin{\phi})\,v_\te + (\sin{\te}\cos{\phi})\,v_\phi\bigr]   \,,  \cr
 P_3 &=&   (\cos{\te})\,v_r-  \bigl(\Cos_\kp(r)\Sin_\kp(r)\bigr)\sin{\te}\,v_\te  \,. 
\end{eqnarray}

\item[(J)]  The three $\kp$-dependent  functions $P_{J1}$, $P_{J2}$, and $P_{J3}$, defined as
$$
  P_{J1} =  i(Y_1)\,\te_L  \,,\quad 
  P_{J2} =  i(Y_2)\,\te_L   \,,\quad 
  P_{J3} =  i(Y_3)\,\te_L   \,, 
$$
that are  given by
\begin{eqnarray}
   P_{J1}  &=& -\Sin_\kp^2(r)\,(\sin\phi\,v_\te + \sin\te\cos\te\cos\phi\,v_\phi) \,,\cr
   P_{J2}  &=&  \Sin_\kp^2(r)\, (\cos\phi\,v_\te - \sin\te\cos\te\sin\phi\,v_\phi) \,,\cr
   P_{J3}  &=&  \Sin_\kp^2(r)\,\sin^2{\te}\,v_\phi   \,.
\end{eqnarray}
\end{itemize}

These Noether momenta satisfy the following property:
\begin {property}
The  Noether  momenta $P_i$ and  $P_{Ji}$, $i=1,2,3$, are constants of the motion for the   geodesic motion.
$$
 \frac{d}{dt} P_i =  \Ga_L(P_i) = 0 \,,{\quad}  \frac{d}{dt} P_{Ji} =  \Ga_L(P_{Ji}) = 0
$$ 
\end{property}
where  $\Ga_L$ denotes the dynamical vector field
$$
  \Ga_L = v_r\,\fracpd{}{r} + v_\te\,\fracpd{}{\te} + v_\phi\,\fracpd{}{\phi} + 
  f_r\,\fracpd{}{v_r} + f_\te\,\fracpd{}{v_\te} + f_\phi\,\fracpd{}{v_\phi}  
$$
with the Lagrangian forces $(f_r,f_\te,f_\phi)$ being  given by 
\begin{eqnarray*}
   f_r  &=& \Cos_\kp(r)\Sin_\kp(r)(v_{\te}^2 + \sin^2\te \, v_\phi^2 ) \,,\cr
   f_\te  &=&    -\Bigl(\frac{2}{\Tan_\kp(r)}\Bigr)v_r v_{\te} +  (\cos\te  \sin\te ) v_\phi^2\,,\cr
   f_\phi  &=&   -2\Bigl(\frac{v_r }{\Tan_\kp(r)} +\frac{v_\te    }{\tan\te } \Bigr)v_\phi\,.
\end{eqnarray*}

\subsection{Hamiltonian formalism}

Under the Legendre transformation the point $(r,\te,\phi,v_r,v_\te,v_\phi)$ of the velocity phase space goes to   the point $ (r,\te,\phi,p_r,p_\te,p_\phi)$ of the phase space  given by
$$
 p_r =v_r  \,,{\quad}  p_\te = \Sin_\kp^2(r)\, v_\te \,,{\quad} p_\phi = \Sin_\kp^2(r)\sin^2\te\, v_\phi
$$
so that the expression of the $\kp$-dependent  Hamiltonian is 
\begin{equation}
 H(\kp) = (\frac{1}{2})\,\Bigl( p_r^2 +  \frac{1}{\Sin_\kp^2(r)} 
 \Bigl(p_\te^2 + \frac{p_\phi^2}{\sin^2\te}\Bigr)\Bigr)  \,, 
\end{equation}
and the images of the six Noether momenta are  
\begin{eqnarray}
  P_1  &=&  (\sin{\te}\cos\phi)\,{p_r} + \Bigl(\frac{\Cos_\kp(r)}{\Sin_\kp(r)}\Bigr)
   \Bigl[(\cos{\te}\cos{\phi})\,p_\te - \bigl(\frac{\sin{\phi}}{\sin\te}\bigr)\,p_\phi\Bigr]     \,,  \cr
 P_2  &=&   (\sin{\te}\sin\phi)\,{p_r} + \Bigl(\frac{\Cos_\kp(r)}{\Sin_\kp(r)}\Bigr)
   \Bigl[(\cos{\te}\sin{\phi})\,p_\te + \bigl(\frac{\cos{\phi}}{\sin\te}\bigr)\,p_\phi\Bigr]      \,,  \cr
 P_3  &=&   (\cos{\te})\,p_r-  \Bigl(\frac{\Cos_\kp(r)}{\Sin_\kp(r)}\Bigr)\sin{\te}\,p_\te  \,,
\label{NmP}
\end{eqnarray}
and
\begin{equation}
   P_{J1}  = -\bigl(\sin\phi\,p_\te +\bigl(\frac{\cos\te}{\sin\te}\bigr)\cos\phi\,p_\phi\bigr)  \,, {\quad}
   P_{J2}  =  \cos\phi\,p_\te - \bigl(\frac{\cos\te}{\sin\te}\bigr)\sin\phi\,p_\phi  \,, {\quad}
   P_{J3} =  p_\phi\,.
\end{equation}
These new six Noether momenta  $P_i$, $P_{Ji}$, $i=1,2,3$, Poisson commute with the Hamiltonian 
$$
 \bigl\{P_i,H\bigr\}= 0 \,,{\quad} \bigl\{P_{Ji},H\bigr\}= 0   \,,{\quad} i=1,2,3, 
$$ 
and the Noether momenta $P_i$ satisfy the Poisson bracket relations
$$
 \bigl\{P_1\,,P_2 \bigr\} = \kp\,P_{J3} \,,{\quad} \bigl\{P_2\,,P_3 \bigr\} = \kp\,P_{J1}
  \,,{\quad}  \bigl\{P_3\,,P_1 \bigr\} =  \kp\,P_{J2}.
$$
We remark the positive sign in these Poisson brackets (in contrast with the negative sign in the Lie brackets). 
The reason is that the Lie bracket between two Hamiltonians vector fields satisfy the property 
$[X_f, X_g] = -\,X_{\{f, g\}}$. That is, the map from the Lie brackets  into the Poisson brackets  ($[\cdot \,,\, \cdot] \to \{\cdot \,,\, \cdot\}$) is linear but introduces a 
change of sign, it is an anti-isomorphism. 

In what follows, as the Noether momenta $P_{Ji}$ coincide with components of the angular momentum, we just write $J_i$ instead of $P_{Ji}$ and we write the other defining Poisson relations as. 
$$
 \bigl\{J_1\,, c_1P_1+c_2P_2+c_3P_3 \bigr\} = c_2 P_3 - c_3  P_2  \,,{\quad} 
 \bigl\{J_2\,, c_1P_1+c_2P_2+c_3P_3 \bigr\} = c_3 P_1 - c_1  P_3\,,
$$
and 
$$
 \bigl\{J_3\,, c_1P_1+c_2P_2+c_3P_3 \bigr\} = c_1 P_2 - c_2  P_1.
$$

 Making use of the expressions of the functions $P_i$, $i=1,2,3$, and of  the angular momenta $J_i$ we obtain  that the sum of their squares take the values  
$$
  P_1^2 + P_2^2 + P_3^2 = p_r^2 + \Bigl( \frac{\Cos_\kp^2(r)}{\Sin_\kp^2(r)} \Bigr) p_\te^2 +
  \Bigl(\frac{\Cos_\kp^2(r)}{\Sin_\kp^2(r)\sin^2\te}\Bigr) p_\phi^2   \,,{\qquad} 
  J_{1}^2 + J_{2}^2 + J_{3}^2 = p_\te^2 +\frac{p_\phi^2}{\sin^2\te}  \,, 
$$
and therefore   the  $\kp$-dependent Hamiltonian can be rewritten as a linear combination of the squares of the six Noether momenta. 
\begin{equation}
 H(\kp) = (\frac{1}{2})\,\Bigl( P_1^2 + P_2^2 + P_3^2 + \kp \,(J_{1}^2 + J_{2}^2 + J_{3}^2)\Bigr)  \,.
\end{equation}

 We close this section mentioning that this property, that is, expressing the Hamiltonian as a function of the Noether momenta (instead of the canonical momenta) is important for the process of quantization of the system;   
see \cite{CRS12Jmp} for the free particle and \cite{CRS17Jpa,CRS19} for general properties of the Killing vector fields and Noether momenta approach to quantization.

\section{Oscillator related Hamiltonian with nonlinear terms $k_1/x_\kp^2$, $k_2/y_\kp^2$, and $k_3/z_\kp^2$}   

In this section we consider  particular examples of Lagrangians of mechanical type where potential terms are added to the geodesic term that correspond to the harmonic oscilator and to the   Smorodinsky-Winternitz system. 

\subsection{The Harmonic Oscillator on the 3-dimensional sphere $S^3$ ($\kp>0)$ and Hyperbolic space $H^3$ ($\kp<0$) }    \label{Sec31}

The following curvature-dependent function represents the Hamiltonian of the isotropic harmonic oscillator  on the spherical  ($\kp>0$), Euclidean, or hyperbolic ($\kp<0$), three-dimensional spaces with constant curvature $\kp$
\begin{equation}
 H(\kp) = \frac{1}{2}\,\Bigl( p_r^2 +  \frac{1}{\Sin_\kp^2(r)}
 \Bigl(p_\te^2 + \frac{p_\phi^2}{\sin^2\te}\Bigr)\Bigr) +  \frac{1}{2}\,\al^2\,\Tan_\kp^2(r) \,, 
\end{equation}
so that, in this way,  the potential of the harmonic oscillator on the unit sphere ($\kp=1$), on the Euclidean space, or on the unit Lobachevsky space ($\kp=-1$) arise as the following three particular cases 
 $$
  V_{1}  = \frac{1}{2}\,\al^2\,\tan^2r   \,,{\qquad} V_{0} = \frac{1}{2}\,\al^2\,r^2\,,{\qquad}
  V_{-1} = \frac{1}{2}\,\al^2\,\tanh^2r  \,. 
$$
The Euclidean oscillator $V_0$, that is a parabolic potential without singularities,  appears in this formalism as making a separation between two different situations (see Fig. 1). 
In the spherical case the potential has an infinite potential barrier on the geodesic circle $r=\pi/(2\sqrt{\kp})$  so that, in this case, the motion is confined on a three-dimensional  half-sphere.
The hyperbolic potential $V_\kp$, $\kp<0$, is a well with finite depth since $\lim{}_{r\to\infty}V_\kp = 1/|\kp|$.

 A quantization of this system was studied in \cite{CRS11Ijtp}  and the  Schr\"odinger equation determined by this Hamiltonian was solved in \cite{CRS12Jpa} (the radial Schr\"odinger equation becomes a $\kp$-dependent Gauss hypergeometric equation that can be considered as a $\kp$-deformation of the confluent hypergeometric equation that appears in the Euclidean case) but making use of another system of coordinates (see Appendix II). 
We also mention that this oscillator  was also studied  in \cite{BallestHerranz09}  using as an approach first  a stereographic projection and then  both Poincar\'e and Beltrami coordinates.

First, in both cases, spherical ($\kp>0$) and hyperbolic ($\kp<0$) spaces, the Hamiltonian system has spherical symmetry  and therefore the three components $(J_1,J_2,J_3)$ of the angular momentum are time preserved in time evolution.

Now let us consider the following three complex functions
$$
 M_{1\kp} = P_1 +  {\ii}\al\, \bigr(\Tan_\kp(r)\bigr)  (\sin\te\cos\phi) \,,{\quad} 
 M_{2\kp} = P_2 +  {\ii}\al\, \bigr(\Tan_\kp(r)\bigr)  (\sin\te\sin\phi)  \,,
$$
$$
 M_{3\kp} = P_3 +  {\ii}\al\, \bigr(\Tan_\kp(r)\bigr)  (\cos\te) \,, 
$$
where $P_i$, $i=1,2,3$, are the Noether momenta given by (\ref{NmP}). 
Then we have
$$
 \bigl\{ M_{1\kp} \,, H(\kp)  \bigr\}  = {\ii}\la_{\kp}\,\al\, M_{1\kp}  \,,{\quad} 
 \bigl\{ M_{2\kp} \,, H(\kp)  \bigr\}  = {\ii}\la_{\kp}\,\al\, M_{2\kp}  \,,{\quad} 
 \bigl\{ M_{3\kp} \,, H(\kp)  \bigr\}  = {\ii}\la_{\kp}\,\al\, M_{3\kp}  \,, 
$$
where $\la_{\kp}$ denotes $\la_{\kp} =1/\Cos_\kp^2(r)$. 
Therefore the moduli  $|\, M_{j\kp}\,|$ of the three functions $M_{j\kp}$, $j=1,2,3$, satisfy 
$$
 \frac{d}{dt}\,|\,M_{1\kp}\,|^2 = \Bigl( \frac{d}{dt}\,M_{1\kp}\Bigr) M_{1\kp}^* 
 +  \,M_{1\kp}\Bigl( \frac{d}{dt}M_{1\kp}^*\Bigr)  
 = ({\ii}\la_{\kp}\al M_{1\kp})M_{1\kp}^*  + M_{1\kp}(- {\ii}\la_{\kp}\al\, M_{1\kp}^*  ) = 0 \,,
$$
$$
  \frac{d}{dt}\,|\,M_{2\kp}\,|^2 = \Bigl( \frac{d}{dt}\,M_{2\kp}\Bigr) M_{2\kp}^* 
  +  \,M_{2\kp}\Bigl( \frac{d}{dt}M_{2\kp}^*\Bigr)  
= ({\ii}\la_{\kp}\al M_{2\kp})M_{2\kp}^*  + M_{2\kp}(- {\ii}\la_{\kp}\al\, M_{2\kp}^*  ) = 0 \,,
$$
$$
  \frac{d}{dt}\,|\, M_{3\kp}\,|^2 = \Bigl( \frac{d}{dt}\,M_{3\kp}\Bigr) M_3^* 
  +  \,M_{3\kp}\Bigl( \frac{d}{dt}M_{3\kp}^*\Bigr)  
= ({\ii}\la_{\kp}\al M_{3\kp})M_{3\kp}^*  + M_{3\kp}(- {\ii}\la_{\kp}\al\, M_{3\kp}^*  ) = 0 \,,
$$
Therefore the following three $\kp$-dependent functions 
\begin{eqnarray}
K_{11\kp} &=&  P_1^2 +  \al^2\, \bigr(\Tan_\kp^2(r)\bigr)  (\sin\te\,\cos\phi)^2  \,,\cr
K_{22\kp} &=&  P_2^2 +  \al^2\, \bigr(\Tan_\kp^2(r)\bigr)  (\sin\te\,\sin\phi)^2   \,,\cr
K_{33\kp} &=&  P_3^2 +  \al^2\, \bigr(\Tan_\kp^2(r)\bigr)  (\cos\te)^2   \,,
\end{eqnarray}
are constants of motion 
$$
 \bigl\{ K_{11\kp}  \,, H(\kp)  \bigr\}  =  0  \,,{\quad} 
 \bigl\{ K_{22\kp}  \,, H(\kp)  \bigr\}  =  0  \,,{\quad} 
 \bigl\{ K_{33\kp}  \,, H(\kp)  \bigr\}  =  0  \,. 
$$
For the same reason we also have that the complex functions $M_{i\kp}M_{j\kp}^*$, $i\ne j$, are constants of motion 
$$
 \bigl\{ M_{1\kp} M_{2\kp}^*  \,, H(\kp)  \bigr\}  =  0  \,,{\quad} 
 \bigl\{ M_{2\kp} M_{3\kp}^*  \,, H(\kp)  \bigr\}  =  0  \,,{\quad} 
 \bigl\{ M_{3\kp} M_{1\kp}^*  \,, H(\kp)  \bigr\}  =  0  \,.  
$$
 If a complex function is a constant of motion for a real Hamiltonian system then it determines two different real constants of motion, its real and imaginary parts:
 $$
 M_{1\kp} M_{2\kp}^*  = K_{12\kp}  +  {\ii}\al \, J_3  \,,{\quad} 
 M_{2\kp} M_{3\kp}^*  = K_{23\kp}  +  {\ii}\al \, J_1  \,,{\quad} 
 M_{3\kp} M_{1\kp}^*  = K_{31\kp}  +  {\ii}\al \, J_2  \,, 
$$
The imaginary parts $\Im(M_{i\kp}M_{j\kp}^*)$,   $i\ne j$,  are related to the components $J_k$, $k=1,2,3$,  of the angular momentum, $\Im(M_{i\kp}M_{j\kp}^*)=\epsilon_{ijk}J_k$.  
 Furthermore,  the three functions $K_{ij\kp}$, such that  $\Re(M_{i}M_j^*)=K_{ij\kp}$,   $i\ne j$,  that are given by
\begin{eqnarray}
K_{12\kp} &=&  P_1 P_2  +  \al^2\, \bigr(\Tan_\kp^2(r)\bigr)   (\sin\te)^2 (\cos\phi\sin\phi)   \,,\cr
K_{23\kp} &=&  P_2 P_3  +  \al^2\, \bigr(\Tan_\kp^2(r)\bigr)   (\sin\te\cos\te) (\sin\phi)   \,,\cr
K_{31\kp} &=&  P_3 P_1  +  \al^2\, \bigr(\Tan_\kp^2(r)\bigr)   (\sin\te\cos\te) (\cos\phi)  \,, 
\end{eqnarray}
Poisson commute with the Hamiltonian
$$
 \bigl\{ K_{12\kp}  \,, H(\kp)  \bigr\}  =  0  \,,{\quad} 
 \bigl\{ K_{23\kp}  \,, H(\kp)  \bigr\}  =  0  \,,{\quad} 
 \bigl\{ K_{31\kp}  \,, H(\kp)  \bigr\}  =  0  \,. 
$$
 These six functions can be considered as the six independent components  of the $\kp$-dependent symmetric Fradkin  matrix
 $$ \left[ K_{ij\kp} \right]  =
\left[\matrix{
 K_{11\kp}  & K_{12\kp}  & K_{13\kp} \cr
 K_{21\kp}  & K_{22\kp}  & K_{23\kp} \cr 
 K_{31\kp}  & K_{32\kp}  & K_{33\kp} \cr  }\right],{\quad} K_{ij\kp} = K_{ji\kp} \,, 
$$
and we can summarize all the Poisson brackets with the  Hamiltonian in a single equation 
$$
 \Bigl\{K_{ij\kp}\,,H(\kp)\Bigr\} = 0 \,,{\quad} i,j =1,2,3. 
$$
Note that the expression of $ K_{ij\kp}$ depends on the Noether momenta instead of on the canonical momenta.
We also note that the functions  $K_{ij\kp} $ are quadratic in the Noether momenta $P_i$, $i=1,2,3$, and as the Noether momenta are linear functions of the canonical momenta $p_a$, $a=r,\te,\phi$, the result is that these functions are also quadratic in the canonical momenta in the given chart. 

The algebraic properties of $\left[ K_{ij\kp} \right]$, that represents the curvature-dependent version of the Fradkin tensor \cite{Fradkin65}, are summarized in the Appendix I.

Another important point is that the Hamiltonian can be rewritten as a sum of the three functions $K_{jj\kp}$, $j=1,2,3$,  and the square of the three angular momenta
\begin{equation}
  H(\kp) = \frac{1}{2}\,\Bigl( K_{11\kp} + K_{22\kp} + K_{33\kp} + \kp \,(J_1^2 + J_2^2 + J_3^2)\Bigr)\,.
\end{equation}
This is an interesting property since it shows that, on spaces of constant curvature, the angular momentum has a direct contribution to the total energy of the system, 
and, as mentioning in the previous section, it is also important in the quantization process  \cite{CRS12Jmp,CRS17Jpa,CRS19}.

The Poisson brackets of these functions are as follows 
$$
 \bigl\{K_{11\kp}  \,,\,  J_1  \bigr\}   =  \bigl\{K_{22\kp}  \,,\,  J_2  \bigr\}   = 
 \bigl\{K_{33\kp}  \,,\,  J_3  \bigr\}   = 0  \,, 
$$
$$
  \bigl\{c_1 K_{11\kp} + c_2 J_1 \,,\, K_{22\kp} + K_{33\kp} + \kp \,(J_2^2 +  J_3^2)  \bigr\}   = 0  \,, 
$$
$$
  \bigl\{c_1 K_{22\kp} + c_2 J_2 \,,\,  K_{11\kp} + K_{33\kp} + \kp \,(J_1^2 + J_3^2)  \bigr\}   = 0  \,, 
$$
$$
  \bigl\{c_1 K_{33\kp} + c_2 J_3 \,,\, K_{11\kp} + K_{22\kp} + \kp \,(J_1^2 + J_2^2)  \bigr\}   = 0  \,, 
$$
where  $c_1$ and $c_2$ are arbitrary  constants. 

We recall that, although  the three components $(J_1,J_2,J_3)$ of the angular momentum do not commute, it is always possible to select a three dimensional Abelian subalgebra generated, for instance, by $H$, $J_1^2 + J_2^2 + J_3^2$ and $J_3$. 
In addition, the above Poisson brackets relations show the existence of other triplets of commuting first integrals as 
$ ( K_{11\kp} , J_1 , K_{22\kp} + K_{33\kp} + \kp \,(J_2^2 + J_3^2) )$, 
$ ( K_{22\kp} , J_2 , K_{11\kp} + K_{33\kp} + \kp \,(J_1^2 + J_3^2) )$, or 
$ ( K_{33\kp} , J_3 , K_{11\kp} + K_{22\kp} + \kp \,(J_1^2 + J_2^2) )$. 

All these results can be  summarized in the following proposition: 
\begin{proposition} \label{Prop1}
 The $\kp$-dependent classical Harmonic Oscillator defined on the 3-dimensional sphere $S^3$ ($\kp>0)$ and on the Hyperbolic space $H^3$ ($\kp<0$) by
$$
 H(\kp) = \frac{1}{2}\,\Bigl( p_r^2 +  \frac{1}{\Sin_\kp^2(r)}
 \Bigl(p_\te^2 + \frac{p_\phi^2}{\sin^2\te}\Bigr)\Bigr) +  \frac{1}{2}\,\al^2\,\Tan_\kp^2(r) 
$$
 is superintegrable with the maximal number of functionally independent constants of motion. 
 It is spherically symmetric so the three components $(J_1,J_2,J_3)$ of the angular momentum are integrals of motion  (this implies  Liouville integrability).  
In addition there is a family of six $\kp$-dependent quadratic functions $K_{ij\kp}$, $i,j =1,2,3$,  that can be considered as the six components of the curvature-dependent version of the symmetric Fradkin tensor.  
\end{proposition}

\subsection{The Smorodinsky-Winternitz (S-W) system  on the 3-dimensional sphere $S^3$ ($\kp>0)$ and on the hyperbolic space $H^3$ ($\kp<0$) }   \label{Sec32}

In what follows  use is made of the following notation
\begin{equation}
 x_\kp = \Sin_\kp(r) \sin \te \cos\phi  \,,{\quad}
 y_\kp = \Sin_\kp(r) \sin \te \sin\phi \,,{\quad}
 z_\kp = \Sin_\kp(r) \cos \te    \,, 
\end{equation}
 such that their  Euclidean limits are
 $$
  \lim{}_{\kp\to 0} (x_\kp, y_\kp, z_\kp) = (r\sin \te \cos\phi , r\sin \te \sin\phi , r\cos \te )
 $$
that correspond to the expression of the Cartesian coordinates $(x,y,z)$ when written in spherical coordinates. 
 We note that the Poisson brackets of these functions with the Noether momenta (\ref{NmP}) are given by
$$
 \bigl\{ x_\kp \,, P_{1} \bigr\}  =  \bigl\{ y_\kp \,, P_{2} \bigr\}  = 
 \bigl\{ z_\kp \,, P_{3} \bigr\}  = \Cos_\kp(r)  \,, 
$$
where we recall that  $ \lim{}_{\kp\to 0}\Cos_{\kp}(r) = 1$. 

 In this section we analyse the following Hamiltonian function which is the spherical ($\kp>0$), Euclidean, or hyperbolic ($\kp<0$),  version of the three-dimensional Smorodinsky-Winternitz (S-W) \cite{Ev90PL} with curvature $\kp$:
\begin{equation}
 H_{SW}(\kp) = \frac{1}{2}\,\Bigl( p_r^2 +  \frac{1}{\Sin_\kp^2(r)}
 \Bigl(p_\te^2 + \frac{p_\phi^2}{\sin^2\te}\Bigr)\Bigr) +  \frac{1}{2}\,\al^2\,\Tan_\kp^2(r) 
 + \Bigl[ \frac{k_1}{x_\kp^2} + \frac{k_2}{y_\kp^2} + \frac{k_3}{z_\kp^2} \Bigr] \,, 
\end{equation}
(the Euclidean version of this system, that is also known as the `caged oscillator' \cite{EvVe08b, KaMil12, ChanuDgRast15,GubLat18}, can be considered as the three-dimensional version of the isotonic oscillator      
\cite{WeisJort79,Zhu87JPa}). 
First, the components $(J_1,J_2,J_3)$ of the angular momentum are not constants of the motion anymore,  but the following three angular momentum related functions
$$
K_{J1} =  J_1^2 +  2 \Bigl( k_2\frac{z_\kp^2}{y_\kp^2} + k_3\frac{y_\kp^2}{z_\kp^2} \Bigr) \,,{\quad}
K_{J2} =  J_2^2 +  2 \Bigl( k_1\frac{z_\kp^2}{x_\kp^2} + k_3\frac{x_\kp^2}{z_\kp^2} \Bigr)  \,,
$$
\begin{equation}
K_{J3} =  J_3^2 +  2 \Bigl( k_1\frac{y_\kp^2}{x_\kp^2} + k_2\frac{x_\kp^2}{y_\kp^2} \Bigr) \,,
\end{equation}
are constants of the motion:
$$
 \bigl\{ K_{J1}  \,, H_{SW}(\kp)  \bigr\}  = 0  \,,{\quad}  
 \bigl\{ K_{J2}  \,, H_{SW}(\kp)  \bigr\}  = 0  \,,{\quad}    
 \bigl\{ K_{J3}  \,, H_{SW}(\kp)  \bigr\}  = 0  \,. 
 $$
 They are   functionally independent, that is 
$  dK_{J1}\,\wedge\, dK_{J2}\,\wedge\, dK_{J3}\ne 0 $,
and  satisfy the following Poisson bracket relations:  
$$
 \bigl\{ K_{J1}\,, K_{J2} + K_{J3}\bigr\} =  0  \,,{\quad}  
 \bigl\{ K_{J2}\,, K_{J1} + K_{J3}\bigr\} =  0  \,,{\quad}  
 \bigl\{ K_{J3}\,, K_{J1} + K_{J2}\bigr\} =  0  \,.
$$
So this system is Liouville integrable (for all the values of $\kp$) with a fundamental set of three integrals of motion 
($H_{SW}(\kp) , K_{Ji}\,, K_{Jj}+K_{Jk}$ ; $i\ne j\ne k$) that Poisson commute. 

On the other hand, the following three functions $K_{jj\kp} $, $j=1,2,3$,  related with the Noether momenta $P_j$, $j=1,2,3$, 
\begin{eqnarray}
 K_{11\kp} &=&  P_1^2 +  \al^2\, \bigr(\Tan_\kp^2(r)\bigr)  (\sin\te\,\cos\phi)^2  + \frac{2 k_1}{(\Tan_\kp(r)\sin \te \cos\phi )^2}  \,, \cr
 K_{22\kp} &=&  P_2^2 +  \al^2\,\bigr(\Tan_\kp^2(r)\bigr)  (\sin\te\,\sin\phi)^2   + \frac{2 k_2}{(\Tan_\kp(r)\sin \te \sin\phi )^2}  \,, \cr
 K_{33\kp} &=&  P_3^2 +  \al^2\,\bigr(\Tan_\kp^2(r)\bigr)  (\cos\te)^2  + \frac{2 k_3}{(\Tan_\kp(r) \cos\te )^2}  \,, 
\end{eqnarray}
are also constants of the motion 
$$
 \bigl\{ K_{11\kp}  \,, H_{SW}(\kp)  \bigr\}   = 0  \,,{\quad} 
 \bigl\{ K_{22\kp}  \,, H_{SW}(\kp)  \bigr\} =  0  \,,{\quad} 
 \bigl\{ K_{33\kp}  \,, H_{SW}(\kp)  \bigr\} =  0  \,. 
$$
Moreover, we have the following two important properties.
First, the following equality is satified
\begin{equation}
 H_{SW}(\kp) = (\frac{1}{2})\,\Bigl( K_{11\kp} + K_{22\kp} + K_{33\kp} 
 + \kp\,\,(K_{J1} + K_{J2} + K_{J3})\Bigr) + \kp(k_1+k_2+k_3) \,. 
\end{equation}
Second, the Poisson brackets among these functions are as follows: 
$$
 \bigl\{K_{11\kp}  \,,\,  K_{J1}   \bigr\}   =  \bigl\{K_{22\kp}  \,,\,  K_{J2}   \bigr\}   = 
 \bigl\{K_{33\kp}  \,,\,  K_{J3}   \bigr\}   = 0  \,, 
$$
$$
  \bigl\{c_1 K_{11\kp} + c_2K_{J1}  \,,\, K_{22\kp} + K_{33\kp} + \kp \,(K_{J2} + K_{J3})  \bigr\}   = 0  \,,
$$
$$
  \bigl\{c_1 K_{22\kp} + c_2K_{J2}   \,,\, K_{11\kp} + K_{33\kp} + \kp \,(K_{J1} + K_{J3})  \bigr\}   = 0  \,, 
$$
$$
  \bigl\{c_1 K_{33\kp} + c_2K_{J3}  \,,\,  K_{11\kp} + K_{22\kp} + \kp \,(K_{J1} + K_{J2})  \bigr\}   = 0  \,,  
$$
where $c_1$ and $c_2$ are arbitrary constants.

Therefore triplets of commuting first integral are, for instance, $ ( K_{11\kp} , K_{J1} , K_{22\kp} + K_{33\kp} + \kp (K_{J2 }+ K_{J3}) )$, 
$ ( K_{22\kp} , K_{J2} , K_{11\kp} + K_{33\kp} + \kp (K_{J1} + K_{J3}) ) $, or $ ( K_{33\kp} , K_{J3} , K_{11\kp} + K_{22\kp} + \kp (K_{J1} + K_{J2}) ) $.

We can summarize all these results in the following proposition:
\begin{proposition} \label{Prop SWkp}   
 The $\kp$-dependent classical Harmonic Oscillator with three additional nonlinear terms defined on the 3-dimensional sphere $S^3$ ($\kp>0)$ and Hyperbolic space $H^3$ ($\kp<0$) 
$$
 H_{SW}(\kp) = \frac{1}{2}\,\Bigl( p_r^2 +  \frac{1}{\Sin_\kp^2(r)}
 \Bigl(p_\te^2 + \frac{p_\phi^2}{\sin^2\te}\Bigr)\Bigr) +  \frac{1}{2}\,\al^2\,\Tan_\kp^2(r) 
 + \Bigl[ \frac{k_1}{x_\kp^2} + \frac{k_2}{y_\kp^2} + \frac{k_3}{z_\kp^2} \Bigr]  \,, 
$$
represents the curvature-dependent version of the 3-dimensional  Smorodinsky-Winternitz system, because
$$
  \lim{}_{\kp\to 0}H_{SW}(\kp) = H_{SW} = \frac{1}{2}\,
  \Bigl( p_r^2 +  \frac{1}{r^2} \Bigl(p_\te^2 + \frac{p_\phi^2}{\sin^2\te}\Bigr)\Bigr) 
 +  \frac{1}{2}\,\al^2\,r^2  + \Bigl[  \frac{k_1}{x^2} + \frac{k_2}{y^2} + \frac{k_3}{z^2} \Bigr] \,. 
$$
It is superintegrable with two sets of three quadratic integrals of motion. 
A first set of three angular  momentum-related functions $K_{Ji}$,  $i=1,2,3$, that are curvature-independent,  and a second set of three $\kp$-dependent functions $K_{ii\kp}$, $i=1,2,3$. 
Two of the functions of  the second set  can be chosen for the total set of five functionally independent integrals of motion. 
\end{proposition}

\subsection{Oscillator 1:1:2  on the 3-dimensional sphere $S^3$ ($\kp>0$) and on the Hyperbolic space $H^3$ ($\kp<0$) with two nonlinear terms of the form $1/x_\kp^2$  and $1/y_\kp^2$  } 

It is well known that the properties of non-central potentials are more complicated to study that those of the central ones (the first negative property is that the angular momentum is not an integral of 
motion). Nevertheless the  oscillator with ratio of frequencies 2:1 appears in  \cite{FrisMSUW65} in the list of superintegrable two-dimensional Euclidean potentials and the three-dimensional oscillator with ratio 2:1:1 also appears in the list of Evans \cite{Ev90PR}. 
The superintegrability of the 2:1 oscillator on the two-dimensional sphere $S^2$ and the hyperbolic plane $H^2$ was study in \cite{RS03Jmp}. Now we consider the three-dimensional system with two additional nonlinear terms.

The following curvature-dependent  function represents the  Hamiltonian of the harmonic oscillator with ratio of frequencies 1:1:2  on the spherical ($\kp>0$), Euclidean, or hyperbolic ($\kp<0$), three-dimensional spaces with constant curvature $\kp$  with two additional $\kp$--dependent nonlinear terms of the form $1/x_\kp^2$  and $1/y_\kp^2$ 
\begin{equation}
 H_{112}(\kp) = \frac{1}{2}\,\Bigl( p_r^2 +  \frac{1}{\Sin_\kp^2(r)}
 \Bigl(p_\te^2 + \frac{p_\phi^2}{\sin^2\te}\Bigr)\Bigr) + V_{112}(\kp) 
 + \frac{k_1}{x_\kp^2} + \frac{k_2}{y_\kp^2},
\end{equation}
where $V_{112}(\kp)$, that denotes the potential of the 1:1:2 oscillator, takes the form 
\begin{equation}
 V_{112}(\kp)  =  \frac{1}{2}\,\al^2\,\frac{1}{1 - \kp(x_\kp^2 + y_\kp^2)}
 \Bigl( x_\kp^2 + y_\kp^2 + 4A_{z\kp}^2 \Bigr)  \,,{\quad} 
 A_{z\kp} = \frac{\Tan_\kp(r)  \cos\te }{1 - \kp\,(\Tan_\kp(r) \cos\te)^2} \,. 
\end{equation}
where the factor $1/(1 - \kp(x_\kp^2 + y_\kp^2))$ and the function $A_{z\kp}$ are obtained as three-dimensional generalizations of  similar functions obtained in \cite{RS02Rmp} in the study of the two-dimensional 1:2 oscillator. 
It is clear that this particular function  satisfies the appropriate Euclidean limit
$$
 \lim{}_{\kp\to 0} \Bigl(V_{112}(\kp)  + \frac{k_1}{x_\kp^2} + \frac{k_2}{y_\kp^2} \Bigr)
 = \frac{1}{2}\,\al^2\,\bigl(  x^2 + y^2 + 4 z^2 \bigr) + \frac{k_1}{x^2} + \frac{k_2}{y^2}  \,. 
$$

The following three quadratic functions are integrals of motion: 
\begin{eqnarray}
  K_{3\kp} &=& P_3^2 + 4 \al^2 A_{z\kp}^2 \,,{\qquad\quad}
  K_{J3}  = J_3^2  +  2 k_2 \Bigl(\frac{x_\kp}{y_\kp}\Bigr)^2  +  2 k_1 \Bigl(\frac{y_\kp}{x_\kp}\Bigr)^2  \,,\cr
  K_{12\kp} &=& (P_1^2 + \kp J_1^2) + (P_2^2 + \kp J_2^2)  +  \al^2  (1 + 4 \kp A_{z\kp}^2) \Bigl(\frac{x_\kp^2+y_\kp^2}{1 - \kp(x_\kp^2+y_\kp^2)}\Bigr)   \cr
 &+& 2 k_2 \Bigl(\frac{1 - \kp x_\kp^2}{y_\kp^2}\Bigr) + 2 k_1 \Bigl(\frac{1 - \kp y_\kp^2}{x_\kp^2}\Bigr)  \,. 
\end{eqnarray}
These three functions  ($K_{3\kp}$, $K_{J3}$, $K_{12\kp}$)  are functionally independent, that is $dK_{3\kp}\,\wedge\, dK_{J3}\,\wedge\, dK_{12\kp}\ne 0$,   and satisfy the following Poisson bracket relations: 
$$
 \bigl\{ K_{3\kp}    \,, H_{112}(\kp) \bigr\}  = 0  \,,{\quad}  
 \bigl\{ K_{J3}  \,, H_{112}(\kp) \bigr\}  = 0  \,,{\quad}  
 \bigl\{ K_{12\kp} \,, H_{112}(\kp)  \bigr\}  = 0  \,, 
$$
which express that they are first integrals and they also Poisson commute among themselves
$$
 \bigl\{ K_{3\kp}    \,, K_{J3} \bigr\}  = 0  \,,{\quad}  
 \bigl\{ K_{J3}  \,, K_{12\kp} \bigr\}  = 0  \,,{\quad}  
 \bigl\{ K_{12\kp}  \,, K_{3\kp}   \bigr\}  = 0  \,,
$$
that is, they generate an Abelian Lie subalgebra; therefore this $\kp$-dependent Hamiltonian is completely integrable in the  Liouville sense.
Moreover,  an interesting property is that the Hamiltonian $H_{112}(\kp)$ can be rewritten as follows: 
 $$  
 H_{112}(\kp) = \frac{1}{2}\,\Bigl( K_{3\kp} + K_{12\kp} + \kp K_{J3} \Bigr)  \,. 
$$

 Furthermore, this $\kp$-dependent system admits two additional integrals of motion of Runge-Lenz type explicitly given by 
\begin{eqnarray}
 K_{RL1} &=& -P_1 J_2  +  \al^2\, \Bigl(\frac{\tan\te \cos\phi}{\Cos_\kp(r)} \Bigr)  A_{z\kp}^2x_\kp 
 - 2  k_1 \Cos_\kp(r) \Bigl(\frac{z_\kp}{x_\kp^2}\Bigr)  \,, \cr
 K_{RL2} &=& P_2 J_1   +  \al^2\, \Bigl(\frac{\tan\te \sin\phi}{\Cos_\kp(r)}  \Bigr)  A_{z\kp}^2 y_\kp 
 - 2  k_2 \Cos_\kp(r)  \Bigl(\frac{z_\kp}{y_\kp^2}\Bigr)  \,,  \label{KRLa}
\end{eqnarray}
that are functionally independent, that is $dK_{RL1}\,\wedge\, dK_{RL2}\ne 0$, as well as  functionally  independent of the other three. 
Therefore this Hamiltonian system possesses five functionally independent integrals of motion, three of them in involution and we can conclude:

\begin{proposition} \label{Prop3}
 The curvature dependent Harmonic Oscillator, with ratio of frequencies 1:1:2 and two additional nonlinear terms of the form $1/x_\kp^2$  and $1/y_\kp^2$, defined on the 3-dimensional sphere $S^3$ ($\kp>0)$ and  on the Hyperbolic space $H^3$ ($\kp<0$) 
$$
 H_{112}(\kp) = \frac{1}{2}\,\Bigl( p_r^2 +  \frac{1}{\Sin_\kp^2(r)}
 \Bigl(p_\te^2 + \frac{p_\phi^2}{\sin^2\te}\Bigr)\Bigr) + V_{112}(\kp) 
 + \frac{k_1}{x_\kp^2} + \frac{k_2}{y_\kp^2},
$$
where $V_{112}(\kp)$ denotes the following $\kp$-dependent potential 
$$
V_{112}(\kp)  =  \frac{1}{2}\,\al^2\,\frac{1}{1 - \kp(x_\kp^2 + y_\kp^2)}
 \Bigl( x_\kp^2 + y_\kp^2 + 4A_{z\kp}^2 \Bigr)  \,,{\quad} 
 A_{z\kp} = \frac{\Tan_\kp(r)  \cos\te }{1 - \kp\,(\Tan_\kp(r) \cos\te)^2},
$$
such that it satisfies the appropriate Euclidean limit
$$
 \lim{}_{\kp\to 0} H_{112}(\kp) =  \frac{1}{2}\,\Bigl( p_r^2 +  \frac{1}{r^2}
 \Bigl(p_\te^2 + \frac{p_\phi^2}{\sin^2\te}\Bigr)\Bigr) + 
  \frac{1}{2}\,\al^2\,\bigl(  x^2 + y^2 + 4 z^2 \bigr) + \frac{k_1}{x^2} + \frac{k_2}{y^2}  \,, 
$$
 is superintegrable with a maximal number of five functionally independent constants of motion. 
It admits three constants of motion ($K_{3\kp}$,  $K_{12\kp}$,  $K_{J3} $) that Poisson commute among them and, in addition,  this system possesses  two $\kp$-dependent quadratic functions $K_{RLj}$, $j=1,2$,  of Runge-Lenz type.  
\end{proposition}

\section{Kepler related Hamiltonian on the 3-dimensional sphere $S^3$ ($\kp>0)$ and on the hyperbolic space $H^3$ ($\kp<0$)  }

Another prototypical example of integrable system is the Kepler problem and therefore we fix in this section our attention on such a problem in the three-dimensional spaces considered in 
the preceding sections.

\subsection{Kepler  Hamiltonian}  \label{Sec41}

The following curvature-dependent  function is the spherical, Euclidean, or hyperbolic, Kepler Hamiltonian with curvature $\kp$ 
\begin{equation}
 H_K(\kp) = \frac{1}{2}\,\Bigl( p_r^2 +  \frac{1}{\Sin_\kp^2(r)}
 \Bigl(p_\te^2 + \frac{p_\phi^2}{\sin^2\te}\Bigr)\Bigr) + \frac{ k}{\Tan_\kp(r)}  \,, 
\end{equation}
i.e., the potentials of the Kepler problem on the unit sphere ($\kp=1$), on the Euclidean space, or on the unit Lobachevsky space ($\kp=-1$), arise as the following three particular cases
 $$
  V_{1}  = \frac{ k}{\tan(r)}  \,,{\qquad} V_{0} = \frac{ k}{r} \,,{\qquad}
  V_{-1} = \frac{ k}{\tanh(r)}.
 $$
The situation is rather similar to the one obtained in Section \ref{Sec31} for the harmonic  oscillator.
Also in this case the Euclidean function $V_0=k/r$  appears in this formalism as making a separation between two different behaviours (see Fig. 2).

 This potential is central (for all the values of $\kp$) so the three components $(J_1,J_2,J_3)$ of the angular momentum are integrals of motion, namely
  $$
 \bigl\{ J_1  \,, H_{K}(\kp)  \bigr\}   = 0   \,,{\quad}  
 \bigl\{ J_2  \,, H_{K}(\kp)  \bigr\}   = 0   \,,{\quad}  
 \bigl\{ J_3  \,, H_{K}(\kp)  \bigr\}   = 0  \,.
$$
and therefore this curvature-dependent system,  as any other central potential,  is Liouville integrable. Moreover,  as in the Euclidean case, there exists an additional set of integrals of motion, because
 the following three functions 
\begin{eqnarray} 
 K_{RL1} &=& (P_2 J_{3}  -  P_3 J_{2}) +  k\, (\sin\te  \cos\phi )  \,,\cr
 K_{RL2} &=& (P_3 J_{1}  -  P_1 J_{3}) +  k\, (\sin\te  \sin\phi )  \,,\cr
 K_{RL3} &=& (P_1 J_{2}  -  P_2 J_{1}) +  k\, (\cos\te )   \,,  \label{KRL}
\end{eqnarray}
that are  functionally independent  
$$   dK_{RL1} \,\wedge\, dK_{RL2} \,\wedge\, dK_{RL3}  \ne 0  \,, 
$$
are integrals of motion 
$$
 \bigl\{ K_{RL1}  \,, H_{K}(\kp)  \bigr\}   = 0   \,,{\quad}  
 \bigl\{ K_{RL2}  \,, H_{K}(\kp)  \bigr\}   = 0   \,,{\quad}  
 \bigl\{ K_{RL3}  \,, H_{K}(\kp)  \bigr\}   = 0  \,.
$$
They must be considered as the curved version of the standard Runge-Lenz vector.
Their Poisson brackets are given by 
$$
 \bigl\{ K_{RL1} \,, K_{RL2}  \bigr\}   = - 2 J_3 \bigl(H_K(\kp) - \kp \,(J_1^2 + J_2 ^2 + J_3^2)\bigr) \,,
$$
$$
 \bigl\{ K_{RL2} \,, K_{RL3}  \bigr\}   = - 2 J_1 \bigl(H_K(\kp) - \kp \,(J_1^2 + J_2 ^2 + J_3^2)\bigr) \,,
$$
$$
 \bigl\{ K_{RL3} \,, K_{RL1}  \bigr\}   = - 2 J_2 \bigl(H_K(\kp) - \kp \,(J_1^2 + J_2 ^2 + J_3^2)\bigr) \,, 
$$
and the Poisson brackets of each one with the angular momenta are  
$$
 \bigl\{ J_{1} \,, c_1K_{RL1} + c_2 K_{RL2} + c_3 K_{RL3} \bigr\}   
 = c_2 K_{RL3} - c_3K_{RL2}  \,,
$$
$$
 \bigl\{ J_{2} \,, c_1K_{RL1} + c_2 K_{RL2} + c_3 K_{RL3} \bigr\}   
 = c_3 K_{RL1} - c_1K_{RL3} \,,
 $$
$$
 \bigl\{ J_{3} \,, c_1K_{RL1} + c_2 K_{RL2} + c_3 K_{RL3} \bigr\}   
 = c_1 K_{RL2} - c_2K_{RL1}  \,, 
 $$
where $c_1$, $c_2$, and $c_3$ are arbitrary constants. 

The preceding results are summarised in the following proposition:
\begin{proposition} \label{Prop4}
 The $\kp$-dependent Kepler Hamiltonian defined in the 3-dimensional sphere $S^3$ ($\kp>0)$ and Hyperbolic space $H^3$ ($\kp<0$) 
$$
 H_K(\kp) = \frac{1}{2}\,\Bigl( p_r^2 +  \frac{1}{\Sin_\kp^2(r)}
 \Bigl(p_\te^2 + \frac{p_\phi^2}{\sin^2\te}\Bigr)\Bigr) + \frac{ k}{\Tan_\kp(r)}
$$
 is superintegrable with very similar properties to those of  the standard  Euclidean Kepler Hamiltonian.
It is spherically symmetric, with the three components ($J_{1}, J_{2} ,J_{3}$)  of the angular momentum as constants of motion, and it also possesses three quadratic constants of motion ($K_{RL1}, K_{RL2} ,K_{RL3}$) representing the components of the curvature-dependent version of the Runge-Lenz vector. 
 \end{proposition}

\subsection{Kepler related Hamiltonian with nonlinear terms $k_1/x_\kp^2$, $k_2/y_\kp^2$, and $k_3/z_\kp^2$}  \label{Sec52}

In this section we will study the following Kepler-related Hamiltonian
\begin{equation}
  H_{K123}(\kp)  = \frac{1}{2})\,\Bigl( p_r^2 +  \frac{1}{\Sin_\kp^2(r)}
 \Bigl(p_\te^2 + \frac{p_\phi^2}{\sin^2\te}\Bigr)\Bigr) + \frac{ k}{\Tan_\kp(r)}
  + \Bigl[   \frac{k_1}{x_\kp^2} + \frac{k_2}{y_\kp^2} + \frac{k_3}{z_\kp^2} \Bigr]    \,,
\end{equation}
where the three additional nonlinear terms,  $k_1/x_\kp^2$,  $k_2/y_\kp^2$  and $k_3/z_\kp^2$,  are just the same as in the S-W system studied in the previous section \ref{Sec32}.

This Hamiltonian admits two different sets of constants of motion. 
A first set is related with the angular momentum and the second set related with the Runge-Lenz vector but of fourth order in the momenta. 

First, the components $(J_1,J_2,J_3)$ of the angular momentum are not integrals of motion anymore but the following three angular momentum related functions$$
  K_{J1}  = J_1^2  +  2 k_2 \Bigl(\frac{z_\kp}{y_\kp}\Bigr)^2  +  2 k_3 \Bigl(\frac{y_\kp}{z_\kp}\Bigr)^2  \,,{\quad} 
  K_{J2}  = J_2^2  +  2 k_1 \Bigl(\frac{z_\kp}{x_\kp}\Bigr)^2  +  2 k_3 \Bigl(\frac{x_\kp}{z_\kp}\Bigr)^2  \,,
$$
\begin{equation}
  K_{J3}  = J_3^2 +  2 k_1 \Bigl(\frac{y_\kp}{x_\kp}\Bigr)^2  +  2 k_2 \Bigl(\frac{x_\kp}{y_\kp}\Bigr)^2  \,, 
\end{equation}
 that are functionally independent, 
$ dK_{J1}\,\wedge\, dK_{J2}\,\wedge\, dK_{J3}\ne 0 $,
Poisson commute with the Hamiltonian, $ \bigl\{ K_{Ji}  \,, H_{K123}(\kp)  \bigr\}   = 0$,
$i=1,2,3$, and satisfy the following Poisson bracket properties  
$$
 \bigl\{ K_{J1}\,, K_{J2} + K_{J3}\bigr\} =  0  \,,{\quad}  
 \bigl\{ K_{J2}\,, K_{J1} + K_{J3}\bigr\} =  0  \,,{\quad}  
 \bigl\{ K_{J3}\,, K_{J1} + K_{J2}\bigr\} =  0  \,.
$$
So this system is Liouville integrable (for all the values of $\kp$) with a fundamental set of three integrals of motion 
($H_{K123}(\kp) , K_{Ji}\,, K_{Jj}+K_{Jk}$ ; $i\ne j\ne k$) that Poisson commute. 

Second, the three  $\kp$-dependent Runge-Lenz functions  (\ref{KRL}) obtained in the previous section, and characterizing to the potential $k/\Tan_\kp(r)$, are no longer integrals of motion. 
Now we prove that this system admits three quartic  constants of motion. 
We will obtain them making use of a method already used in  \cite{CRS21Jpa}  in the study superintegrable systems on three-dimensional conformally Euclidean spaces and that it is related with the existence  of certain complex functions.

Let us now denote by $R_{i\kp}$, $i=1,2,3$,  the following Runge-Lenz-related functions 
\begin{eqnarray}
 R_{1\kp} &=& K_{RL1} + 2 \bigl( \Cos_\kp(r)\Sin_\kp(r)\bigr) ( \sin\te \cos\phi ) 
 \,\Bigl( \frac{k_1}{x_\kp^2}  + \frac{k_2}{y_\kp^2} + \frac{k_3}{z_\kp^2} \Bigr)  \,, \cr 
 R_{2\kp} &=& K_{RL2} +2 \bigl( \Cos_\kp(r)\Sin_\kp(r)\bigr)  (\sin\te \sin\phi ) 
  \,\Bigl( \frac{k_1}{x_\kp^2}  + \frac{k_2}{y_\kp^2} + \frac{k_3}{z_\kp^2} \Bigr)  \,, \cr 
 R_{3\kp} &=& K_{RL3} +2 \bigl( \Cos_\kp(r)\Sin_\kp(r)\bigr)  (\cos\te ) 
 \,\Bigl( \frac{k_1}{x_\kp^2}  + \frac{k_2}{y_\kp^2} + \frac{k_3}{z_\kp^2} \Bigr)  \,,  
\end{eqnarray}
where the quadratic functions $K_{RLi}$, $i=1,2,3$, were  defined in the previous section.

In fact, in the particular case $(k\ne 0,k_1=k_2=k_3=0)$, these three functions reduce to three components of the Runge-Lenz vector. 
These functions are not (in the general case)  integrals of motion but when one of the additional terms is not present  then the corresponding function $R_j$ becomes an integral of motion. 
That is, we have the following property 
$$
{\rm If} {\quad}   k_j=0  \quad {\rm then}\quad  \bigl\{ R'_{j\kp}  \,, H_{K123}'(\kp)  \bigr\}   = 0   \,, 
 $$
where $R'_{j\kp}$,   and $H_{K123}'(\kp)$ denote the function $R_{j\kp}$ and  the Hamiltonian $H_{K123}(\kp)$, respectively,  but without the $k_j$-term.

Let us first remark that the functions $x_\kp$, $y_\kp$, $z_\kp$, and the Noether momenta $P_i$, $i=1,2,3$, satisfy the following relation 
$$
  x_\kp P_1 + y_\kp P_2 + z_\kp P_3 = p_r\Sin_\kp(r)  \,.
$$
Then the three  functions $R_{j\kp}$, $ j=1,2,3$, and the three $\kp$-dependent functions
$$ 
 \bigl(p_r\Sin_\kp(r)\bigr)/x_\kp \,,{\quad}     \bigl(p_r\Sin_\kp(r)\bigr)/y_\kp \,,{\quad}   
 \bigl(p_r\Sin_\kp(r)\bigr)/z_\kp  \,,
$$
are related among them by the time derivatives. 
More precisely,  we have
$$
  \bigl\{ R_{1\kp}\,, H_{K123} \bigr\} =  -\, 2 k_1 \lambda_{1\kp}\,\frac{1}{x_\kp} \bigl(p_r\Sin_\kp(r)\bigr) 
  \,,{\quad}   \Bigl\{\frac{1}{x_\kp} \bigl(p_r\Sin_\kp(r)\bigr) \,, H_{K123} \Bigr\} 
 = \lambda_{1\kp}\,R_{1\kp} \,, 
$$
$$
  \bigl\{R_{2\kp}\,, H_{K123} \bigr\} = -\,2 k_2 \lambda_{2\kp}\,\frac{1}{y_\kp} \bigl(p_r\Sin_\kp(r)\bigr ) 
  \,,{\quad}   \Bigl\{\frac{1}{y_\kp} \Bigl(p_r\Sin_\kp(r)\bigr) \,, H_{K123} \Bigr\}  
 =   \lambda_{2\kp}\,R_{2\kp} \,,
$$
$$
  \bigl\{ R_{3\kp}\,, H_{K123} \bigr\} =  -\, 2 k_3 \lambda_{3\kp}\,\frac{1}{z_\kp} \bigl(p_r\Sin_\kp(r)\bigr) 
  \,,{\quad}  \Bigl\{\frac{1}{z_\kp} \Bigl(p_r\Sin_\kp(r)\bigr) \,, H_{K123} \Bigr\}  
  =  \lambda_{3\kp}\,R_{3\kp} \,,
$$
where the coefficients $\la_{j\kp }$, $j=1,2,3$, take the forms
$$
  \lambda_{1\kp} = \frac{1}{x_\kp^2} \,,{\quad}  \lambda_{2\kp} = \frac{1}{y_\kp^2} \,,{\quad}
  \lambda_{3\kp} = \frac{1}{z_\kp^2} \,. 
$$
The properties of these functions are stated in the following proposition
\begin{proposition} \label{Prop5}
Let   $N_{j\kp}$, $j=1,2,3$, denote  the following complex functions  
$$
 N_{1\kp} = R_{1\kp} + {\ii} \sqrt{2 k_1}\,\Bigl(\frac{ p_r\Sin_\kp(r)}{x_\kp}\Bigr) \,,{\quad}
 N_{2\kp} = R_{2\kp} + {\ii} \sqrt{2 k_2}\,\Bigl(\frac{ p_r\Sin_\kp(r)}{y_\kp}\Bigr) \,,{\quad}
 N_{3\kp} = R_{3\kp} + {\ii} \sqrt{2 k_3}\,\Bigl(\frac{ p_r\Sin_\kp(r)}{z_\kp}\Bigr) \,,
$$
Then the time derivatives of  these functions satisfy the following  relations
$$
  \frac{d}{dt}\,N_{1\kp} = -\, {\ii}\sqrt{2 k_1}\, \lambda_{1\kp}\,N_{1\kp} \,,{\quad}
  \frac{d}{dt}\,N_{2\kp} = -\, {\ii}\sqrt{2 k_2}\, \lambda_{2\kp}\,N_{2\kp}  \,,{\quad}
  \frac{d}{dt}\,N_{3\kp} = -\, {\ii}\sqrt{2 k_3}\, \lambda_{3\kp}\,N_{3\kp}  \,.
$$
\end{proposition} 
Therefore the moduli  $|\, N_{j\kp}\,|$ of the functions $N_{j\kp}$, $j=1,2,3$, satisfy 
$$
  \frac{d}{dt}\,|\, N_{1\kp}\,|^2 = \Bigl( \frac{d}{dt}\,N_{1\kp}\Bigr)N_{1\kp}^* 
  +  N_{1\kp}   \Bigl( \frac{d}{dt}N_{1\kp}^*\Bigr)    =  (-\, {\ii} \sqrt{2 k_1}\, \lambda_{1\kp} 
  + {\ii}\sqrt{2 k_1}\, \lambda_{1\kp} ) \Bigl(N_{1\kp}N_{1\kp}^*\Bigr) = 0 \,,
$$
$$
  \frac{d}{dt}\,|\,N_{2\kp} \,|^2 = \Bigl( \frac{d}{dt}\,N_{2\kp} \Bigr) N_{2\kp}^* 
  +  N_{2\kp}    \Bigl( \frac{d}{dt}N_{2\kp} ^*\Bigr)     =   (-\, {\ii} \sqrt{2 k_2}\, \lambda_{2\kp} 
  + {\ii} \sqrt{2 k_2}\, \lambda_{2\kp} ) \Bigl(N_{2\kp} N_{2\kp}^*\Bigr) = 0 \,,
$$
$$
  \frac{d}{dt}\,|\, N_{3\kp}\,|^2 = \Bigl( \frac{d}{dt}\,N_{3\kp}\Bigr) N_3^* 
 +  N_{3\kp}    \Bigl( \frac{d}{dt}N_{3\kp}^*\Bigr)  = (-\, {\ii} \sqrt{2 k_3}\, \lambda_{3\kp}  
 + {\ii} \sqrt{2 k_3}\, \lambda_{3\kp} ) \Bigl(N_{3\kp}N_{3\kp}^*\Bigr) = 0  \,. 
$$
Hence the three functions  $K_{Rj}$,  $j=1,2,3$, given by 
$$
 K_{R1} = |\,N_{1\kp}\,|^2  = R_{1\kp}^2 + {2k_1}\, \Bigl(\frac{ p_r\Sin_\kp(r)}{x_\kp}\Bigr)^2  \,,{\quad}
 K_{R2} = |\,N_{2\kp}\,|^2  = R_{2\kp}^2 + {2k_2}\, \Bigl(\frac{ p_r\Sin_\kp(r)}{y_\kp}\Bigr)^2  \,,{\quad} 
$$
\begin{equation}
K_{R3} = |\,  N_{3\kp}\,|^2  = R_{3\kp}^2 + {2k_3}\, \Bigl(\frac{p_r\Sin_\kp(r)}{z_\kp}\Bigr)^2 \,, 
\end{equation}
are quartic  constants of motion:
$$
  \bigl\{ K_{Rj}\,, H_{K123}(\kp)\bigr\} =  0 \,,{\qquad} j=1,2,3.
  $$
  The preceding result can be summarised in  the following proposition: 
\begin{proposition} \label{Prop KNL}
 The $\kp$-dependent Kepler Hamiltonian with three additional nonlinear terms, $k_1/x_\kp^2$, $k_2/y_\kp^2$, and $k_3/z_\kp^2$,  defined on the 3-dimensional sphere $S^3$ ($\kp>0)$ 
 and on ghe hyperbolic space $H^3$ ($\kp<0$) 
$$
  H_{K123}(\kp)  = \frac{1}{2}\,\Bigl( p_r^2 +  \frac{1}{\Sin_\kp^2(r)}
 \Bigl(p_\te^2 + \frac{p_\phi^2}{\sin^2\te}\Bigr)\Bigr) + \frac{ k}{\Tan_\kp(r)}
  + \Bigl[   \frac{k_1}{x_\kp^2} + \frac{k_2}{y_\kp^2} + \frac{k_3}{z_\kp^2}  \Bigr],
$$
with Euclidean limit 
$$
  \lim{}_{\kp\to 0}H_{K123}(\kp)  = \frac{1}{2}\,\Bigl( p_r^2 +  \frac{1}{r^2}
 \Bigl(p_\te^2 + \frac{p_\phi^2}{\sin^2\te}\Bigr)\Bigr) + \frac{ k}{r}
 + \Bigl[  \frac{k_1}{x^2} + \frac{k_2}{y^2} + \frac{k_3}{z^2} \Bigr],
$$
 is maximally superintegrable with a first set of three  angular-momentum-related quadratic constants of motion  $(K_{J1}, K_{J2}, K_{J3})$ and a second set $(K_{R1},K_{R2},K_{R3})$ of three curvature-dependent constants of motion of fourth order in the momenta.
\end{proposition}

\section{Final comments } 

We have studied the superintegrability of Hamiltonian systems defined on three-dimensional configuration spaces of constant curvature. 
As observed in the Introduction the two more important superintegrable systems are just the harmonic oscillator and the Kepler problem and because of this we have focused our study on these two systems (usually known as Bertrand potentials) as well as to some other related systems obtained from them by addition of nonlinear terms.

There are certain important points that are fundamental for the approach presented in this article. 
We mention two. First, we have presented a curvature-dependent formalism (all the functions depend of $\kp$ as a parameter) but, given a superintegrable Euclidean system, then many different $\kp$-dependent potentials can be constructed with the same flat limit; the important point is that  if we require that the superintegrability must be preserved then this condition singles out  a very particular system among all the possible curved version of the Euclidean system.
Second, the curvature-dependent formalism we have presented permit us the study of the Hamiltonian system at the same time in both curved manifolds; that is,  spherical (curvature $\kp$ positive) and hyperbolic (curvature $\kp$ negative). 
This is also a very important point since these two spaces are geometrically rather different but, in spite of this, first the Hamiltonian function $H(\kp)$ and then all the integrals of motion $K_{j\kp}$ can be expressed in an unique form valid for the two spaces.

We have proved the quadratic superintegrability (and we have obtained all the integrals of motion) of three oscillators;  the  isotropic harmonic, the Smorodinsky-Winternitz (S-W) system, and the 2:1:1 oscillator with nonlinear terms 
\begin{itemize}
\item The Harmonic Oscillator on the 3-dimensional sphere $S^3$ ($\kp>0)$ and Hyperbolic space $H^3$ ($\kp<0$)
$$
 H(\kp) = \frac{1}{2}\,\Bigl( p_r^2 +  \frac{1}{\Sin_\kp^2(r)}
 \Bigl(p_\te^2 + \frac{p_\phi^2}{\sin^2\te}\Bigr)\Bigr) +  \frac{1}{2}\,\al^2\,\Tan_\kp^2(r) \,. 
$$
$$
 \lim_{\kp\to 0}H(\kp) = \frac{1}{2}\,\Bigl( p_r^2 +  \frac{1}{r^2}
 \Bigl(p_\te^2 + \frac{p_\phi^2}{\sin^2\te}\Bigr)\Bigr) +  \frac{1}{2}\,\al^2\,r^2 \,.
$$
\item Isotropic harmonic oscillator with additional terms of the form $k_1/x_\kp^2$, $k_2/y_\kp^2$, and $k_3/z_\kp^2$:
$$
 H_{SW}(\kp) = \frac{1}{2}\,\Bigl( p_r^2 +  \frac{1}{\Sin_\kp^2(r)}
 \Bigl(p_\te^2 + \frac{p_\phi^2}{\sin^2\te}\Bigr)\Bigr) +  \frac{1}{2}\,\al^2\,\Tan_\kp^2(r) 
 + \Bigl[ \frac{k_1}{x_\kp^2} + \frac{k_2}{y_\kp^2} + \frac{k_3}{z_\kp^2} \Bigr]  \,.
$$
$$
  \lim{}_{\kp\to 0}H_{SW}(\kp) =  \frac{1}{2}\,  \Bigl( p_r^2 +  \frac{1}{r^2} \Bigl(p_\te^2 
  + \frac{p_\phi^2}{\sin^2\te}\Bigr)\Bigr) 
 +  \frac{1}{2}\,\al^2\,r^2  + \Bigl[  \frac{k_1}{x^2} + \frac{k_2}{y^2} + \frac{k_3}{z^2} \Bigr]   \,. 
$$
\item  Oscillator 1:1:2  on the 3-dimensional sphere $S^3$ ($\kp>0)$ and on the hyperbolic space $H^3$ ($\kp<0$) with two nonlinear terms of the form $1/x_\kp^2$  and $1/y_\kp^2$: 
$$
 H_{112}(\kp) = \frac{1}{2}\,\Bigl( p_r^2 +  \frac{1}{\Sin_\kp^2(r)}
 \Bigl(p_\te^2 + \frac{p_\phi^2}{\sin^2\te}\Bigr)\Bigr) + \frac{1}{2}\,\al^2\, 
 \Bigl(\frac{x_\kp^2 + y_\kp^2 + 4A_{z\kp}^2}{1 - \kp(x_\kp^2 + y_\kp^2)}\Bigr)  
 +  \Bigl[\frac{k_1}{x_\kp^2} + \frac{k_2}{y_\kp^2} \Bigr]   \,.
$$
$$
  \lim{}_{\kp\to 0}H_{112}(\kp) =   \frac{1}{2}\,  \Bigl( p_r^2 +  \frac{1}{r^2} \Bigl(p_\te^2 
  + \frac{p_\phi^2}{\sin^2\te}\Bigr)\Bigr) 
 +   \frac{1}{2}\,\al^2\,\bigl(  x^2 + y^2 + 4z^2 \bigr) 
 + \Bigl[\frac{k_1}{x^2} + \frac{k_2}{y^2}  \Bigr]   \,. 
$$
\end{itemize}
and we have also proved the quadratic superintegrability of the Kepler problem with curvature $\kp$ and the higher-order superintegrability of the Kepler problem with additional nonlinear terms:
\begin{itemize}
\item Kepler Hamiltonian on the 3-dimensional sphere $S^3$ ($\kp>0)$ and on the hyperbolic space $H^3$ ($\kp<0$)
$$
 H_K(\kp) = \frac{1}{2}\,\Bigl( p_r^2 +  \frac{1}{\Sin_\kp^2(r)}
 \Bigl(p_\te^2 + \frac{p_\phi^2}{\sin^2\te}\Bigr)\Bigr) + \frac{ k}{\Tan_\kp(r)}   \,.
$$
\item Curvature-dependent Kepler Hamiltonian with three additional terms of the form $k_1/x_\kp^2$, $k_2/y_\kp^2$, and $k_3/z_\kp^2$
$$
  H_{K123}(\kp)  = \frac{1}{2}\,\Bigl( p_r^2 +  \frac{1}{\Sin_\kp^2(r)}
 \Bigl(p_\te^2 + \frac{p_\phi^2}{\sin^2\te}\Bigr)\Bigr) + \frac{ k}{\Tan_\kp(r)}
  + \Bigl[ \frac{k_1}{x_\kp^2} + \frac{k_2}{y_\kp^2} + \frac{k_3}{z_\kp^2} \Bigr]   \,.
$$
$$
  \lim{}_{\kp\to 0}H_{K123}(\kp) =   \frac{1}{2}\,  \Bigl( p_r^2 +  \frac{1}{r^2} \Bigl(p_\te^2 
  + \frac{p_\phi^2}{\sin^2\te}\Bigr)\Bigr) 
+ \frac{ k}{r}  + \Bigl[ \frac{k_1}{x^2} + \frac{k_2}{y^2} + \frac{k_3}{z^2} \Bigr]   \,.
$$
\end{itemize}

We finalise with the following two comments.
(i) In the two cases, oscillator and Kepler, the existence of several integrals of motion appears as related with the properties of certain complex functions (functions $M_{j\kp}$ for the oscillator in Section \ref{Sec31} and functions $N_{j\kp}$ for Kepler in Section \ref{Sec52}). 
In fact, a very similar situation was already obtained in  ref. \cite{CRS21Jpa} in the study of superintegrable Hamiltonian systems  on 3-dimensional  conformally Euclidean spaces.  
A natural question is if this complex-related method is limited to these two particular systems or it can be applied to other different Hamiltonian systems. 
(ii) The  study of classical superintegrability can also be considered  as a first step for the study of the corresponding quantum versions (we recall that quantum superintegrability is related with the degeneracy of the energy levels as in the hydrogen atom). 
The behaviour of the functions $M_{j\kp}$ and $N_{j\kp}$  shows a certain relation with the properties of classical ladder functions studied in \cite{DelisleHussinKN}. 
 An interesting point is if the quantization of the functions  $M_{j\kp}$ and $N_{j\kp}$ as appropriate operators  can be related with quantum ladder operators. 
Thus, quantum version of the properties presented in this paper can also be considered as a matter to be studied.

\section{Appendix I. Properties of the matrix  $[K_{ij\kp}]$}

The symmetric matrix $\left[ K_{ij\kp} \right]$ of the $\kp$-depending integrals of motion,  
$ \bigl\{K_{ij\kp}\,,H(\kp)\bigr\} = 0 $, obtained in the section (\ref{Sec31}) represents   a generalization of the Fradkin tensor \cite{Fradkin65} for the dynamics of the curvature-dependent Hamiltonian $H(\kp)$. Now we present its more important algebraic properties 

\begin{itemize}
\item[(i)]  The trace of the matrix $[K_{ij\kp}]$, that in the Euclidean case is just the Hamiltonian, is now the Hamiltonian plus a curvature-dependent term (related with the angular momentum) that vanish in the Euclidean limit
 $$
 \tr[K_{ij\kp}] = K_{11\kp} + K_{22\kp} + K_{33\kp} \,,{\quad}
\tr[K_{ij\kp}]  + \kp \,(J_1^2 + J_2^2 + J_3^2) = 2 H(\kp)  \,. 
$$

\item[(ii)]  The matrix $[K_{ij\kp}]$ is singular, that is, 
$\det[K_{ij\kp}] = 0$. 
In fact, the six matrix elements $K_{ij\kp}=K_{ji\kp}$, $i,j=1,2,3$,  are six different integrals of motion  for the Hamiltonian $H(\kp) $ and, as only five of them can be independent, the equation $\det[K_{ij}] = 0$ states an algebraic relation between them.

\item[(iii)]  The action of $\left[ K_{ij} \right]$ on the angular momentum is given by 
$$ 
\left[\matrix{
 K_{11\kp}  & K_{12\kp}  & K_{13\kp} \cr
 K_{21\kp}  & K_{22\kp}  & K_{23\kp} \cr 
 K_{31\kp}  & K_{32\kp}  & K_{33\kp} \cr  }\right] 
 \left[\matrix{J_{1} \cr J_{2} \cr J_{3} }\right] =0 ,
$$
that can be rewritten as the following three equations 
$$
 K_{11\kp} J_1 + K_{12\kp} J_2 + K_{13\kp} J_3 = 0   \,,
$$
$$
 K_{21\kp} J_1 + K_{22\kp} J_2 + K_{23\kp} J_3 = 0   \,,
$$
$$
 K_{31\kp} J_1 + K_{32\kp} J_2 + K_{33\kp} J_3 = 0   \,. 
 $$
The contraction of $[K_{ij\kp}]$ with the angular momentum gives zero.

\item[(iv)]  The following relations between the components of the matrix are true: 
$$
x_\kp^2 K_{22\kp} - 2 x_\kp y_\kp K_{12\kp} + y_\kp^2 K_{11\kp} =  (\Cos_{\kp}(r))^2J_{3}^2  \,,{\qquad}
y_\kp^2 K_{33\kp} - 2 y_\kp z_\kp K_{23\kp} + z_\kp^2 K_{22\kp} =  (\Cos_{\kp}(r))^2J_{1}^2  \,, 
$$
$$ z_\kp^2 K_{11\kp} - 2 z_\kp x_\kp K_{31\kp} + x_\kp^2 K_{33\kp} =  (\Cos_{\kp}(r))^2J_{2}^2 ,
$$
where we recall that $ \lim{}_{\kp\to 0} \Cos_{\kp}(r) = 1$.

\item[(v)]  The following relations between the components of the matrix are true: 
$$
 K_{11\kp}  K_{22\kp}  - K_{12\kp}^2 = \al^2 J_3^2  \,,{\quad}
 K_{22\kp}  K_{33\kp}  - K_{23\kp}^2 = \al^2 J_1^2  \,,{\quad}
 K_{33\kp}  K_{11\kp}  - K_{31\kp}^2 = \al^2 J_2^2  \,.
$$
\item[(vi)]  The following three algebraic properties are true 
\begin{eqnarray*}
K_{ij\kp} x_{i\kp} x_{j\kp}  &=&  2 \bigl(x_\kp^2 + y_\kp^2 + z_\kp^2\bigr) H(\kp)  - (J_{1}^2 + J_{2}^2 + J_{3}^2)   \,,\cr 
 K_{ij\kp} x_{i\kp}P_j &=&  (p_r\Sin_\kp(r)) \Bigl(2 H(\kp) -  \kp  (J_1^2 + J_{2}^2 + J_{3}^2) \Bigr)  \,,\cr 
 K_{ij\kp} P_i P_j &=&  (P_1^2+P_2^2+P_3^2)^2 + \al^2   (\Tan_{\kp}(r))^2\,p_r^2   \,,\cr 
\end{eqnarray*}
where we have made use of the following equality
$$
  x_\kp P_1 + y_\kp P_2 + z_\kp P_3 = p_r\Sin_\kp(r) \,. 
$$
\end{itemize}

\section{Appendix II. Two other alternative approaches  }

We have studied the dynamics on the curvature constant spaces $S^3$ and $H^3$ by making use of the curvature-dependent trigonometric and hyperbolic functions $\Cos_{\kp}(x) $ and $\Sin_{\kp}(x) $. 
In this way the expression of the differential element of distance $ds^2(\kp) $  and the  kinetic function $ T_g(\kp) $ 
take the following form when written in geodesic polar coordinates $(r,\te,\phi)$
$$
  ds^2(\kp) =  dr^2  +  \Sin_\kp^2(r)\,d\te^2   +  \Sin_\kp^2(r)\,\sin^2\te d\phi^2   \,,   
$$
and
$$
 L = T_g(\kp)  = (\frac{1}{2})\,\left(v_r^2 + \Sin_\kp^2(r)\, v_\te^2 +  \Sin_\kp^2(r)\sin^2\te \,v_\phi^2 \right) \,. 
$$

Next we present two other different approaches that can be obtained from this one by making use of a change of the geodesic  distance $r$ but preserving the angular coordinates. 

\begin{enumerate}
\item Let us consider the $\kp$-dependent change $(r,\te,\phi) \to (\rho,\te,\phi)$ given by
$\rho = \Sin_\kp(r) $.
Then, when written in these new coordinates, the $\kp$-dependent metric and  kinetic term become
$$
 ds_{\kp}^2 = \frac{dr^2}{1 - \kp \, \rho^2} + \rho^2\,d\te^2  + \rho^2\sin^2 \te\,d\phi^2  \,,
$$
and
$$
 L = T_g(\kp)  = (\frac{1}{2})\,\Bigl(\frac{v_\rho^2}{1 - \kp\,\rho^2} + \rho^2\,v_{\te}^2
 + \rho^2\sin^2 \te\,v_{\phi}^2\Bigr)    \,.  
$$
In this case, the Lagrangians of the harmonic oscillator and Kepler problem take the form
$$
 L(\kp)   = (\frac{1}{2})\,\Bigl(\frac{v_\rho^2}{1 - \kp\,\rho^2} + \rho^2\,v_{\te}^2
 + \rho^2\sin^2 \te\,v_{\phi}^2\Bigr)  - \frac{\al^2}{2}\left(\frac{\rho^2}{1 - \kp\,\rho^2} \right)  \,, 
$$
and
$$
 L (\kp)  = (\frac{1}{2})\,\Bigl(\frac{v_\rho^2}{1 - \kp\,\rho^2} + \rho^2\,v_{\te}^2
 + \rho^2\sin^2 \te\,v_{\phi}^2\Bigr)  - k\left(\frac{\sqrt{1 - \kp\,\rho^2}}{\rho} \right)  \,. 
$$
This formalism was used for example in \cite{ CRS11Jmp,CRS12Jmp,CRS11Ijtp,CRS12Jpa}
\item Let us now consider a new  $\kp$-dependent change $(r,\te,\phi) \to (R,\te,\phi)$ given by
$R = \Tan_\kp(r)$.
Then, when written in these new coordinates, the $\kp$-dependent metric and  kinetic term become
$$
 ds_{\kp}^2 = \frac{d\rho^2}{(1 + \kp\, R^2)^2} + \frac{R^2\,d\te^2}{(1 + \kp\, R^2)}   +
\frac{\rho^2}{(1 + \kp\, R^2)}  \sin^2 \te\,d\phi^2  \,,
$$
and 
$$
 L = T_g(\kp)  = (\frac{1}{2})\,\Bigl(\frac{v_R^2}{(1 + \kp\, R^2)^2} + \frac{R^2\,v_\te^2}{(1 + \kp\, R^2)}   +
\frac{R^2\sin^2 \te}{(1 + \kp\, R^2)}  \,v_\phi^2 \Bigr)
$$
In this case, the Lagrangians of the harmonic oscillator and Kepler problem take the form
$$
 L(\kp)  = (\frac{1}{2})\,\Bigl(\frac{v_R^2}{(1 + \kp\, R^2)^2} + \frac{R^2\,v_\te^2}{(1 + \kp\, R^2)}   +
\frac{R^2\sin^2 \te}{(1 + \kp\, R^2)}  \,v_\phi^2 \Bigr) - \frac{\al^2}{2}R^2  \,,
$$
and
$$
 L(\kp)  = (\frac{1}{2})\,\Bigl(\frac{v_R^2}{(1 + \kp\, R^2)^2} + \frac{R^2\,v_\te^2}{(1 + \kp\, R^2)}   +
\frac{R^2\sin^2 \te}{(1 + \kp\, R^2)}  \,v_\phi^2 \Bigr) - \frac{k}{R}  \,. 
$$

This approach is the one studied by Higgs in Ref.
\cite{Higgs79} (the study of Higgs was originally limited to a
spherical geometry but the idea can be extended to the
hyperbolic space).

\end{enumerate}

We note that  both radial variables, $\rho$ and $R$, are well defined. 
In the hyperbolic $\kp<0$ case the two functions $\Sin_\kp(r)$ and $\Tan_\kp(r)$ are
positive for $r>0$ and concerning the spherical $\kp>0$ case this
property is also true because then $r$ is restricted to a bounded
interval.

 The situation can be summarised as follows.
We have obtained three alternative ways of describing the Lagrangian/Hamiltonian systems on spaces of constant curvature: the original trigonometric/hyperbolic formalism 
 and the two other approaches  obtained from it. 
 Of course, each one of these three different approaches has its own 
 characteristics and advantages.

\section*{Acknowledgments}

J.F.C. and M.F.R. acknowledge support from research Projects No. PGC2018-098265-B-C31 (MINECO, Madrid)  and DGA-E48/20R (DGA, Zaragoza),  
M.S. acknowledges support by the research Projects No.  VA137G18 and BU229P18 (Junta de Castilla y Le\'on).


\begin{figure}\centerline{
\includegraphics{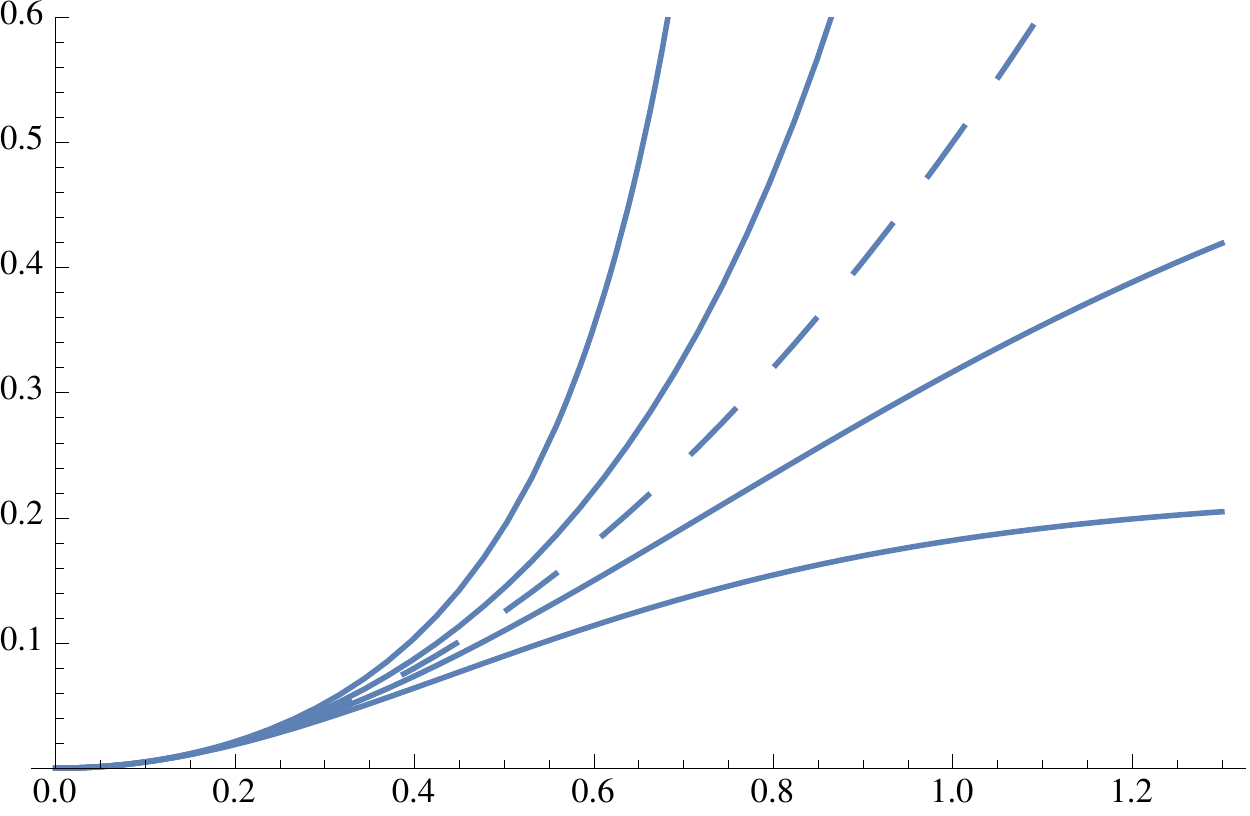}   }

\caption{Plot of the potential $V_\kp$, $\al=1$, as a function of $r$, for $\kp< 0$ (lower curves), $\kp = 0$ (dashed line) and $\kp > 0$ (upper curves).}
\label{Fig1}
\end{figure}

\begin{figure}\centerline{
\includegraphics{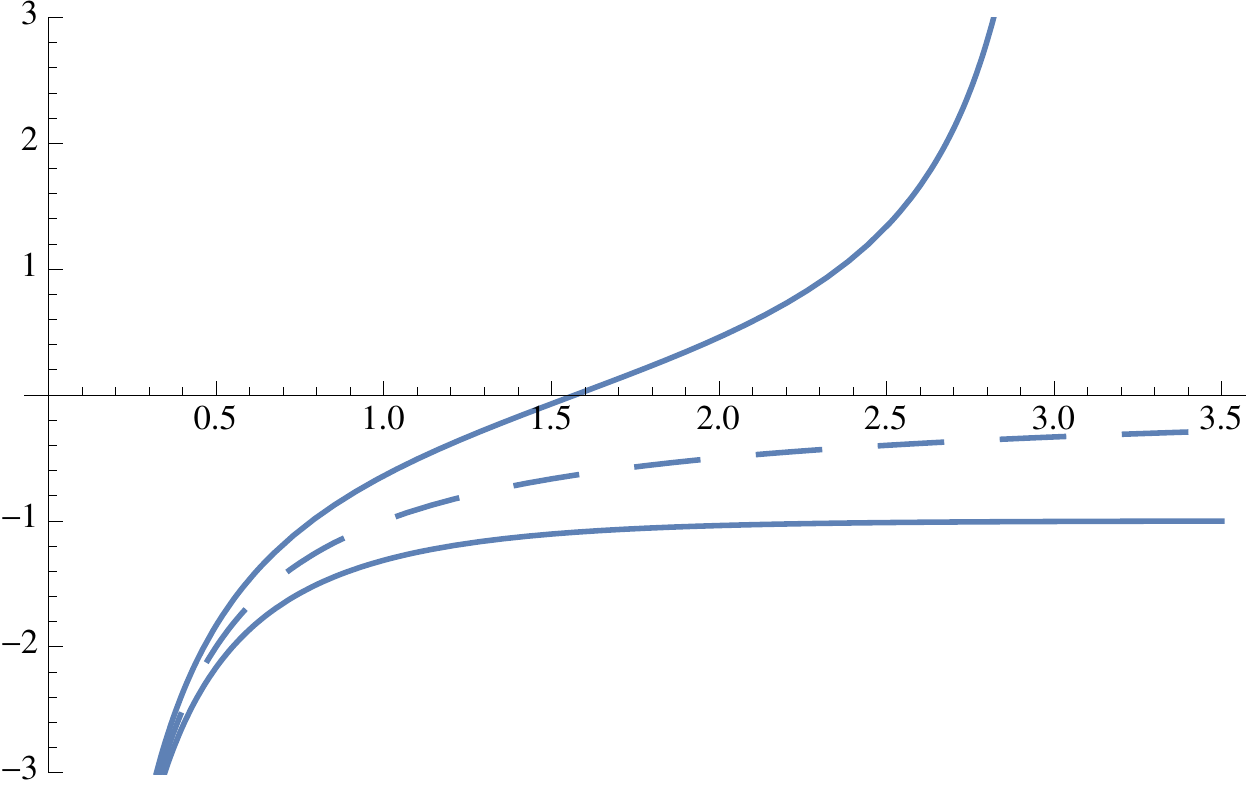}    }

\caption{ Plot of the Kepler potential $-1/\Tan_\kp(r)$ as a function of $r$, for the unit sphere  $\kp=1$  (upper curve), Euclidean potential  (dashed line), and unit Lobachewski space  $\kp=-1$  (lower curve). The three functions are singular at $r=0$ but the Euclidean function $V_0$  appears in this formalism as making a separation between two different behaviours. In fact $V_0$ is the only potential that vanishes at long distances.} 
\label{Fig2}
\end{figure}

  \newpage
{\small
   }

\end{document}